\documentclass[
reprint,
superscriptaddress,
nofootinbib,
 amsmath,amssymb,
 aps,
 onecolumn,
floatfix,
]{revtex4-2}

\usepackage{graphicx}% Include figure files
\usepackage{dcolumn}% Align table columns on decimal point
\usepackage{bm}% bold math
\usepackage{longtable}

\usepackage{xcolor,ulem}

\begin{document}

\title{An Exponential Reduction in Training Data Sizes for Machine Learning Derived Entanglement Witnesses} 

\author{Aiden R. Rosebush}
\affiliation{
Dept of Electrical \& Computer Engineering, University of Toronto, Toronto, Ontario, Canada M5S 3G4
}

\author{Alexander C. B. Greenwood}
\affiliation{
Dept of Electrical \& Computer Engineering, University of Toronto, Toronto, Ontario, Canada M5S 3G4
}

\author{Brian T. Kirby}
\affiliation{DEVCOM Army Research Laboratory, Adelphi, MD 20783 USA}
\affiliation{Tulane University, New Orleans, LA 70118 USA}

\author{Li Qian}
\affiliation{
Dept of Electrical \& Computer Engineering, University of Toronto, Toronto, Ontario, Canada M5S 3G4
}

\date{\today}

\begin{abstract} 
We propose a support vector machine (SVM) based approach for generating an entanglement witness that requires exponentially less training data than previously proposed methods. SVMs generate hyperplanes represented by a weighted sum of expectation values of local observables whose coefficients are optimized to sum to a positive number for all separable states and a negative number for as many entangled states as possible near a specific target state. Previous SVM-based approaches for entanglement witness generation used large amounts of randomly generated separable states to perform training, a task with considerable computational overhead. Here, we propose a method for orienting the witness hyperplane using only the significantly smaller set of states consisting of the eigenstates of the generalized Pauli matrices and a set of entangled states near the target entangled states. With the orientation of the witness hyperplane set by the SVM, we tune the plane's placement using a differential program that ensures perfect classification accuracy on a limited test set as well as maximal noise tolerance. For \(N\) qubits, the SVM portion of this approach requires only \(O(6^N)\) training states, whereas an existing method needs \(O(2^{4^N})\). We use this method to construct witnesses of 4 and 5 qubit GHZ states with coefficients agreeing with stabilizer formalism witnesses to within 6.5 percent and 1 percent, respectively. We also use the same training states to generate novel 4 and 5 qubit W state witnesses. Finally, we computationally verify these witnesses on small test sets and propose methods for further verification.

%Finally, we propose methods for physical and computational verification of these witnesses.%
\end{abstract}

\maketitle

\section{\label{sec:intro} Introduction}
Quantum entanglement is a key resource exploited in
emerging communication \cite{PhysRevLett.98.060503}, \cite{zhong}, imaging \cite{chen}, and computing
\cite{lukens} technologies. These technologies are starting to rely on higher
dimensional quantum systems, meaning groups of many quantum 2-level
systems (qubits) or many-level systems (qudits)\cite{thomas,kues,imany}. 
Hence, the proper implementation of near-term quantum technologies requires the ability to accurately and efficiently characterize quantum states of many-dimensional systems.

One conventional approach to characterizing quantum systems is through quantum state tomography (QST), which involves measuring the expectation values of the system for a complete set of observables, typically Pauli operators or Positive Operator-Valued Measures (POVMs). However, this process is resource-intensive. For a system comprising N particles, each in a d-dimensional Hilbert space, QST necessitates $O(d^{2N})$ measurements. This exponential increase in measurements, corresponding to the degrees of freedom in the system, renders the process impractical for large systems \cite{thew}.

Efforts to improve the resource scaling of full state tomography are ongoing. These include modifications to conventional state reconstruction approaches, such as those based on maximum-likelihood estimation \cite{dfvj} or Bayesian methods \cite{lukens2}, and the incorporation of deep learning techniques \cite{lohani, danaci, lohani_2023}. Recently, adaptations of transformer models, originally developed for large language processing, have shown promise in enhancing the efficiency of QST, albeit within a limited scope and still requiring a significant number of measurements \cite{transformers}.

Rather than performing full QST it is also possible to extract key features of a quantum state from a subset of measurements. Recent methods, such as neural networks \cite{ma} and convex hull approximations \cite{lu}, aim to reduce the number of necessary measurements, though their efficiency relative to the size of the system can vary. In particular, the neural network approach still required \(O(3^N)\) measurements for general n-qubit, few body systems where the entanglement is partially known, and only showed a significant reduction in measurements in the case where the states were shown to be GHZ-type \cite{ma}. The convex hull method effectively relies on the results of at least \(10^3\) binary classifications for two qubit states, which often does not result in a reduction of the necessary number of measurements \cite{lu}. 

For specific tasks like classifying a state as entangled or separable, constructing an entanglement witness $\mathcal{W}$ offers a more measurement-efficient method. The expectation value $\langle \mathcal{W}\rangle$ is non-negative for all separable states and negative for certain entangled states. This approach, which does not require reconstruction of the complete density matrix, typically needs only $O(d^{N})$ measurements. This scaling is a result of the claim that each witness \(W\) will have no more than \((d + 1)^N\) features, as particle of dimension d will have at most (d+1) mutually-unbiased bases, with each basis corresponding to its own generalized Pauli matrix \(\sigma_k\) \cite{PhysRevApplied.19.034058}. While entanglement witnesses using the stabilizer formalism \cite{toth} and fidelity method \cite{guhne} have been developed, the number of measurements required can still be substantial, depending on the specific context \cite{guhne}.

A recent study has proposed an innovative approach to generate entanglement witnesses by utilizing support vector machines (SVMs), a technique commonly employed in machine learning for binary classification problems \cite{PhysRevApplied.19.034058}. SVMs work by constructing hyperplanes that optimally divide two data classes in a vector space, maximizing the margin between the hyperplane and the nearest data points of each class. This concept bears a resemblance to linear entanglement witnesses in quantum physics which are formulated as a linear combination of local observables, essentially measurements applied to unknown quantum states. The integration of SVMs in this context is particularly intriguing as it suggests a novel method for optimizing entanglement witnesses, such as the systematic reduction in the number of features while bounding the inference error and the resulting derivation of a W state witness that required fewer measurements than standard fidelity witnesses.

The SVM-based approach to deriving entanglement witnesses suffers from two main drawbacks. The first is the scaling properties of the required training set with the dimension of the Hilbert space. The methods introduced in \cite{PhysRevApplied.19.034058} generated training sets by randomly sampling states, and hence, sampling states with a constant density requires exponentially more samples as a function of dimension. The second significant issue is the result of a critical difference between SVMs and witnesses in that the latter aims to optimize the noise tolerance of entangled states only, meaning we would want to maximize the margin on only one side of the hyperplane and make the hyperplane as close to the boundary of the separable states as possible unlike with SVMs that attempt to maximize the margin on both sides equally.

Here, we introduce methods for addressing the training data scaling and the optimal placement of the hyperplane issues in the previous study.
In particular, we introduce a method of sampling training data using the eigenstates of the generalized Pauli observables, combined with a technique using gradient descent to maximize the noise tolerance of witnesses we derive and ensure they correctly classify all separable states. We estimate that our approach scales as \(O(6^N)\) for \(N\) qubit witnesses as compared with the scaling of \(O(2^{4^N})\) found in \cite{PhysRevApplied.19.034058}. Further, we account for the default symmetric placement of the hyperplane by an SVM by searching for the separable state which is most misclassified  and using it to adjust the bias term for the hyperplane of the witness generated with the SVM method. Finally, we also use an algorithm called recursive feature elimination (RFE) to find witnesses with fewer terms and which require fewer measurements. The RFE algorithm repeatedly removes the features of the witness which have the most negligible impact on the noise tolerance and retrains the model. It stops when a specified number of features remain or when the noise tolerance drops below some specified value. Using these techniques, we propose several new 4-qubit W-state witnesses, which require far fewer measurements than full state tomography, and numerically verify that each witness we generate correctly classifies a large sample of separable and entangled training data near the targets.

The rest of the paper is organized as follows. Section II details the scheme for generating witnesses with SVMs as outlined in \cite{PhysRevApplied.19.034058}. Section III introduces our reduced training dataset, and Section IV describes a method of correcting witnesses to optimize noise tolerance and ensure perfect accuracy. In section \ref{sec:conf} we use our new method to confirm 4 and 5 qubit GHZ witnesses, and in section \ref{sec:rfe} we extend this method to the generation of new 4 and 5 qubit W state witnesses. In section \ref{sec:numerical_verification} we discuss our method for numerically verifying the new W state witnesses. 
After the conclusion in section \ref{sec:conclusion}, the appendices cover technical details of our method and the full results. In Appendix \ref{sec:gradient_descent}, we provide a brief overview of gradient descent and what frameworks we used to write the differential program introduced in section \ref{sec:mso}. In Appendix \ref{sec:mso_parameterization}, we define how we parameterized separable mixed states which that differential program optimizes. In Appendix \ref{sec:perms}, all the permutations of 4 and 5 qubit pure k-separable states are listed, which we used in the parameterization of the optimizer. Finally in Appendix \ref{sec:coeff}, we present the normalized coefficients of our proposed 4 and 5 qubit W state witnesses.

\section{\label{sec:old_scheme} Original SVM Training Scheme}

\subsection{\label{sec:old_scheme_SVMs} SVMs}

In machine learning, Support Vector Machines (SVMs) are employed for binary classification tasks. SVMs operate by finding a hyperplane in a feature space that separates two data classes. This hyperplane, also known as a decision boundary, is chosen to maximize the margin, which is the distance between each class's nearest points and the hyperplane itself \cite{svm_textbook}. This concept is visually depicted in Figure \ref{fig: SVM}, which illustrates how data points on either side of the hyperplane are classified into two distinct categories.

An SVM assigns data points to categories based on a set of coefficients that form a weighted sum of the feature coordinates of each data point, along with a bias term. If this weighted sum, when calculated for a particular data point, is positive, the SVM assigns a +1 label to that point, indicating it belongs to one category. A negative sum results in a -1 label, placing the point in the other category.

\subsection{\label{sec:old_scheme_Feature_Space} Feature Space}
In the context of this work, feature space refers to the complete set of expectation values of the generalized Pauli operators for states which can be decomposed into qubits. The generalized Pauli operators we use here can be written as \(\sigma_k = \sigma_{k_1} \otimes \sigma_{k_2} \otimes \dots \otimes \sigma_{k_N}\) for an \(N\) qubit system, with \(\sigma_{k_i} \in \{I, X, Y, Z\}\), and where \(k\) represents a string of Pauli operator labels, such as \(XYZ\).
In an $N$-qubit Hilbert space, the set of all possible Kronecker products of single-qubit Pauli operators, including the identity, comprises $4^N$ distinct elements and constitutes a complete orthogonal basis. Feature space represents the real-valued space of all expectation values of a given set of operators, which for the Pauli operators is $4^N$ dimensional.
Quantum states can be encoded into feature vectors in this basis, which we denote as \(f\), where each element of \(f\) corresponds to a different expectation value of a generalized Pauli observable.

Since the generalized Pauli operators are orthogonal, and our witness is composed of a linear combination of generalized Pauli operators, it also corresponds to a hyperplane in this feature space. Therefore, we need to make a connection between feature space and Hilbert space. With a complete tomographic basis, the process of measurement is invertible \cite{thew} and so there must exist a one-to-one mapping between Hilbert space and feature space.

Furthermore, the sets of all separable states are by definition convex in Hilbert space, and we always choose a convex set of entangled training data as discussed in section \ref{sec:newdata}. It is easy to show both sets are convex in feature space as well. Suppose that \(\rho\) is some entangled or separable state, which can be written as a convex combination of other entangled or separable states as \(\rho = \sum_{i} p_i\rho_i\), with each \(p_i \in [0,1]\) and \(\sum_i p_i = 1\). Element \(f_k\) of the vector in feature space corresponding to \(\rho\) can be written as 

\begin{equation}
\label{eq: feature_def}
    f_k = Tr(\sigma_k \rho).
\end{equation}

 We could also write 

 \begin{equation}
 \label{eq: feature_sum}
  f_k = Tr(\sigma_k  \sum_{i} p_i\rho_i),
 \end{equation}
 
 which by linearity can be rewritten as 

 \begin{equation}
 \label{eq: feature_rewrite}
 f_k = \sum_{i} p_i Tr(\sigma_k \rho_i),
 \end{equation}
 
 which is a convex combination of the feature vectors corresponding to each $\rho_i$. Hence if a set of states is closed under convex combinations in Hilbert space then it must also be closed under convex combinations in feature space.

In words, since any convex combination of separable states is still separable, then any convex combination of the feature vectors of separable states must correspond to the feature vector of some other separable state. The same reasoning applies to our convex subset of entangled states as defined in section \ref{sec:newdata}. This proof is clearly reversible, in the sense that any class of states whose feature vectors form convex set must also be a convex set in Hilbert space.

Notably, the contrapositive of this argument must also be true: For a set of states in Hilbert space which is not convex, then the set of corresponding feature vectors must also not be convex. In the case of fully separable pure states, which are not closed under convex combinations, we can see the corresponding set of feature vectors does not appear convex after the uniform sampling done to produce the red shape in Figure \ref{fig: e_o}.

As noted in \ref{sec: old_method_SVM_witnesses} and defined by \ref{eq: noise}, we choose entangled training states such that every convex combination also produces an entangled state within the set, meaning that our entangled training states are disjoint from the set of separable states in feature space. Since separable and certain entangled feature vectors form convex sets, they must also be disjoint, as otherwise we could find some separable feature vector which can be written as a convex combination of the entangled feature vectors we have chosen despite the one-to-one mapping to Hilbert space where the sets are disjoint.  We must be able to find a hyperplane which acts as a boundary between them, as a result of the Hahn-Banach theorem \cite{Boyd}. Since we can always find such a boundary, we should always be able to construct an entanglement witness for a given target entangled state \cite{PhysRevApplied.19.034058}. The conceptual geometry of entangled and separable states in feature space is illustrated in Figure \ref{fig: SVM}.

We denote the generalized Pauli observables by removing the ``\(\otimes\)'' tensor product symbol, so for example we would refer to the three qubit operator \(X \otimes X \otimes X\) as \(XXX\). Since representations of quantum states in Hilbert space must be normalized, the identity term (for three qubits, \(III\)) always has an expectation value of 1, so the coefficient for the identity term effectively is the bias term for the hyperplane. For the rest of the paper, we will refer to the identity term as the bias term. In section \ref{sec:mso} we will discuss how the coefficient of the bias term can be tuned to maximize the noise tolerance of the derived witness by changing the distance between the plane and the nearest separable state.

The Pauli and generalized Pauli operators, whose corresponding feature vectors $f$, are a tomographically complete basis, e.g., given $f$ we can infer the complete density matrix of the corresponding quantum state. Hence, the goal of an SVM-derived entanglement witness is to find subsets of $f$ that allow for the construction of a hyperplane in feature space that is capable of classifying the states of interest. As long as the hyperplane requires fewer than $4^N$ (for $N$ qubits) elements of $f$ it is more efficient than full-state tomography. In section \ref{sec:rfe} we will consider recursive feature elimination methods with the explicit goal of finding the shortest length feature vectors required for a given classification problem.

\subsection{\label{sec: old_method_SVM_witnesses} SVM Derived Witnesses}

\begin{figure*}
\includegraphics[width=5.5in,height=2.5in]{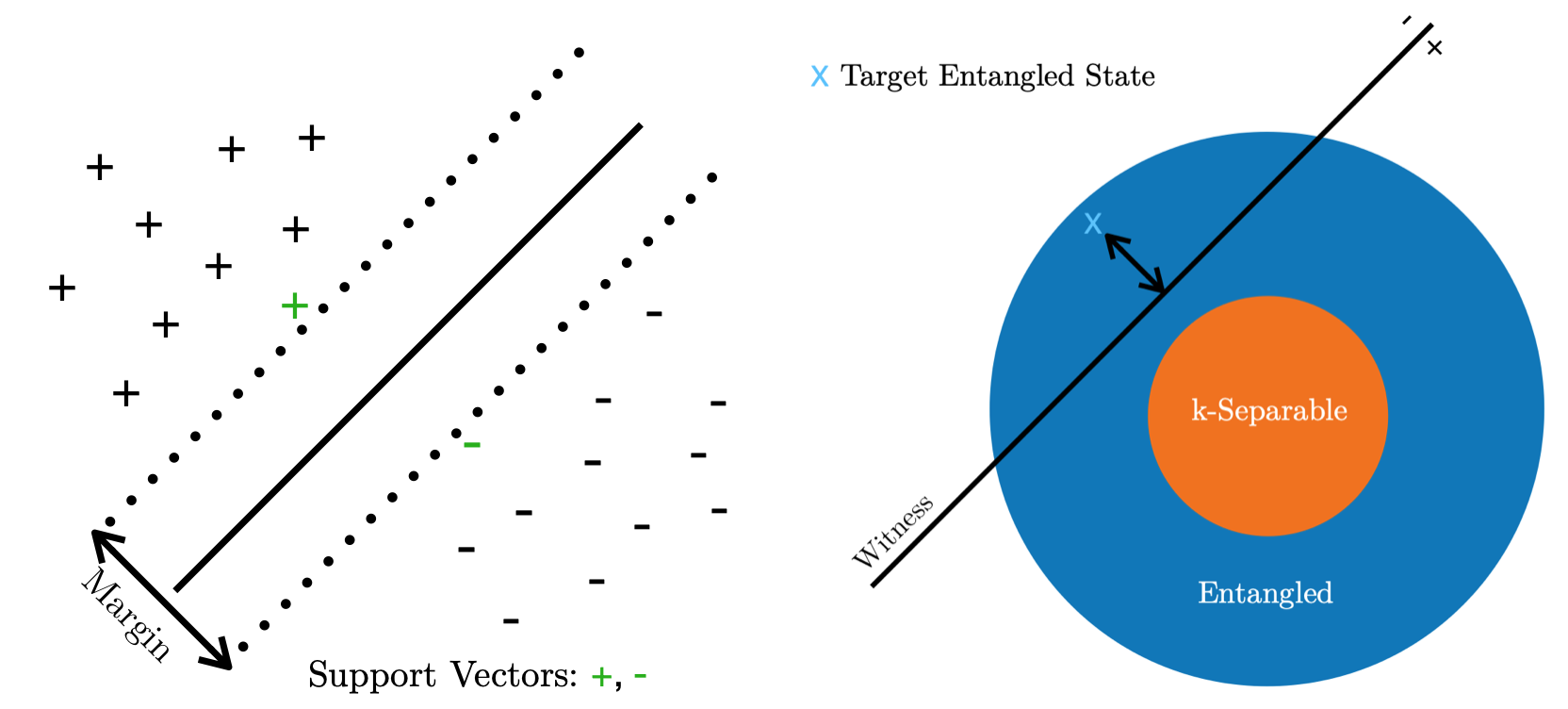}
\caption{Left: The boundary hyperplane and margin a SVM finds between two categories of data (in this case + and -). The support vectors are shown in green. Right: An entanglement witness hyperplane separating the target entangled state, denoted by the light blue X, and the separable states, denoted by the red region. The noise tolerance, depicted by the arrow, of the entanglement witness can be adjusted by adjusting the bias term.}
\label{fig: SVM}
\end{figure*}

In the context of quantum physics, enhancing noise tolerance in an entanglement witness is analogous to optimizing the margin in a binary classification problem. The objective is to maintain a negative expectation value for the witness across a broad range of entangled states. Unlike SVMs, which optimize margins on both sides of a hyperplane, our goal is to maximize the distance between the hyperplane and the target entangled state, while being as close as possible to the nearest separable state. This ensures the decision boundary abuts the boundary of separable states. Ideally, a suitable witness would identify at least one separable state with an exact expectation value of zero but none with a value below zero. Conversely, the aim is for a maximum number of entangled states to exhibit an expectation value below zero. This concept can be visualized by comparing an SVM-derived decision boundary with an entanglement witness, as shown in Figure \ref{fig: SVM}.

The asymmetric nature of an entanglement witness as compared to a standard SVM means that adjustment to the bias term is necessary and must be made exact to ensure the highest noise tolerance while avoiding misclassifying any separable states. In the original paper for SVM derived witnesses \cite{PhysRevApplied.19.034058}, the bias term was adjusted by subtracting from it the lowest expectation value of the separable states sampled for training to ensure no separable states could have a negative value with the final witness. Adjusting this way further required a dense sampling of feature space and significant computational load, as the expectation values of all the training data would had to be found the minimum identified, meaning the runtime of the original method scales with the training data size. In this paper, we will introduce a new method, described in sections \ref{sec:newdata} and \ref{sec:mso}, for adjusting the bias term of the witness found from a smaller set of training states using a differential program. In particular, the differential program will optimize over the entire set of k-separable states to find the separable state with the lowest expectation value and subtract this from the bias term to guarantee that the resulting witness does not misclassify any separable states. Note that classifications with an asymmetrical margin can also be addressed with methods like cost sensitive SVMs \cite{cost_SVM} and weighted SVMs \cite{weight_SVM}, where different classes of data are assigned different losses for a given margin found by the learning model. In our case we use the ordinary SVM training scheme first and then adjust the bias term separately.

Given a training data set in feature space, and labels expressing the correct sign of the expectation value of the witness, the witness can be trained using an SVM. Training allows the program to learn which linear combination of features will assign the same sign labels as the training states.

To detect as many entangled states near the target as possible, it is desirable for an entanglement witness to have as large a noise tolerance as possible. Here, we define noise tolerance as the largest value of \(p\in [0,1]\) for which
\begin{equation} \label{eq: noise} \rho = (1-p)\rho_e + p\frac{\mathbb{I}}{2^N}\end{equation}
satisfies
\begin{equation} \label{eq: entangled_label} Tr(\rho \mathcal{W}) < 0\end{equation}  
for some target entangled state \(\rho_e\) in a system of \(N\)
qubits \cite{PhysRevApplied.19.034058}. The state given in \ref{eq: noise} in general defines what is known as a Werner state \cite{taming}.

SVMs were used successfully in \cite{PhysRevApplied.19.034058} to find the optimal GHZ state witness and several W state witnesses with various numbers of features. However, training data size for the SVMs in \cite{PhysRevApplied.19.034058} is large, which included 500,000 pure biseparable states and 2 million entangled states. The latter were sampled for training using (\ref{eq: noise}), with \( \rho_e = |W\rangle \langle W|\) for a W state witness or \(\rho_e = |GHZ\rangle \langle GHZ| \) for a GHZ state witness, and \( p \in [0,0.25]\).

\section{\label{sec:newdata} New Training Data and Method}

In this section we consider a method for reducing the number of data points required for obtaining an SVM-derived witness as compared to that in section \ref{sec:old_scheme} (and \cite{PhysRevApplied.19.034058}).
To begin we note that, in general, only data points close to the boundary hyperplane between classes need to be used for training.
These are called ``Support
Vectors'' \cite{svm_textbook}, and are illustrated in Figure \ref{fig: SVM} as the green points nearest the separating plane. Engineering training sets to make training more efficient by focusing on the most important examples is an area of active research in machine learning \cite{lohani2022data}, and reducing training data sizes specifically for SVMs has been studied in other contexts  \cite{Wang, Nan}.

Here we propose using the eigenstates of the generalized Pauli operators as a reduced training set for finding classification hyperplanes with the correct orientation for SVM-derived witnesses. We later adjust the location of this plane via a differential program.
These eigenstates are extreme points of the set of separable states in feature space and can be easily parameterized.
They have the useful property that they correspond to extreme expectation values (-1 and 1) for all generalized Pauli operators. Since support vectors are necessarily extreme points of separable states in feature space on the boundary with entangled states, and we have chosen to construct witnesses with linear combinations of Pauli operators, these eigenstates should provide support vectors for classification hyperplanes with the correct orientation for every possible target entangled state. Additionally, since the Pauli operators form a complete set the training set obtained from them samples all orientations of the convex volume formed by them. For this reason the same separable training data can be used to generate an approximate witness for any target. These states are defined below, and the resulting size of training data is compared to the original approach \cite{PhysRevApplied.19.034058}.

For entangled training states we use Werner states, defined in equation (\ref{eq: noise}), which we produce by varying \(p\) incrementally between 0 and 0.25, chosen to much lower than the known limit of 0.526 for \(p\) \cite{taming} where the the resulting state remains entangled. For arbitrary states where this limit is unknown, a smaller limit \(p\) could be used, or even just the target entangled state itself. Here we used \(10^3\) such points, instead of \(2 \times 10^6\) as was the case with 3 qubits. After substantial fine tuning of the hyperparameters in the SVM algorithm, including learning rate, batch size, and regularization, we were later able to produce similar results with only a single entangled training state with \(p = 0.25\), but the small increase in training data allowed us to generate our best results with much less tuning work.

In our method, a SVM is first trained using the fully separable eigenstates and these \(10^3\) entangled states, which we show results in correct coefficients for each expectation value except that of the bias term. To fix the bias term, the separable state with the lowest expectation value is found using gradient descent in a technique called differential mixed state optimization, instead of finding the lowest expectation value of all the randomly sampled separable states in the original paper \cite{PhysRevApplied.19.034058}.

\subsection{\label{sec:eigenstates} Eigenstates of Generalized Pauli Matrices}

\begin{figure*}
\includegraphics[width=5.32153in,height=3.7in]{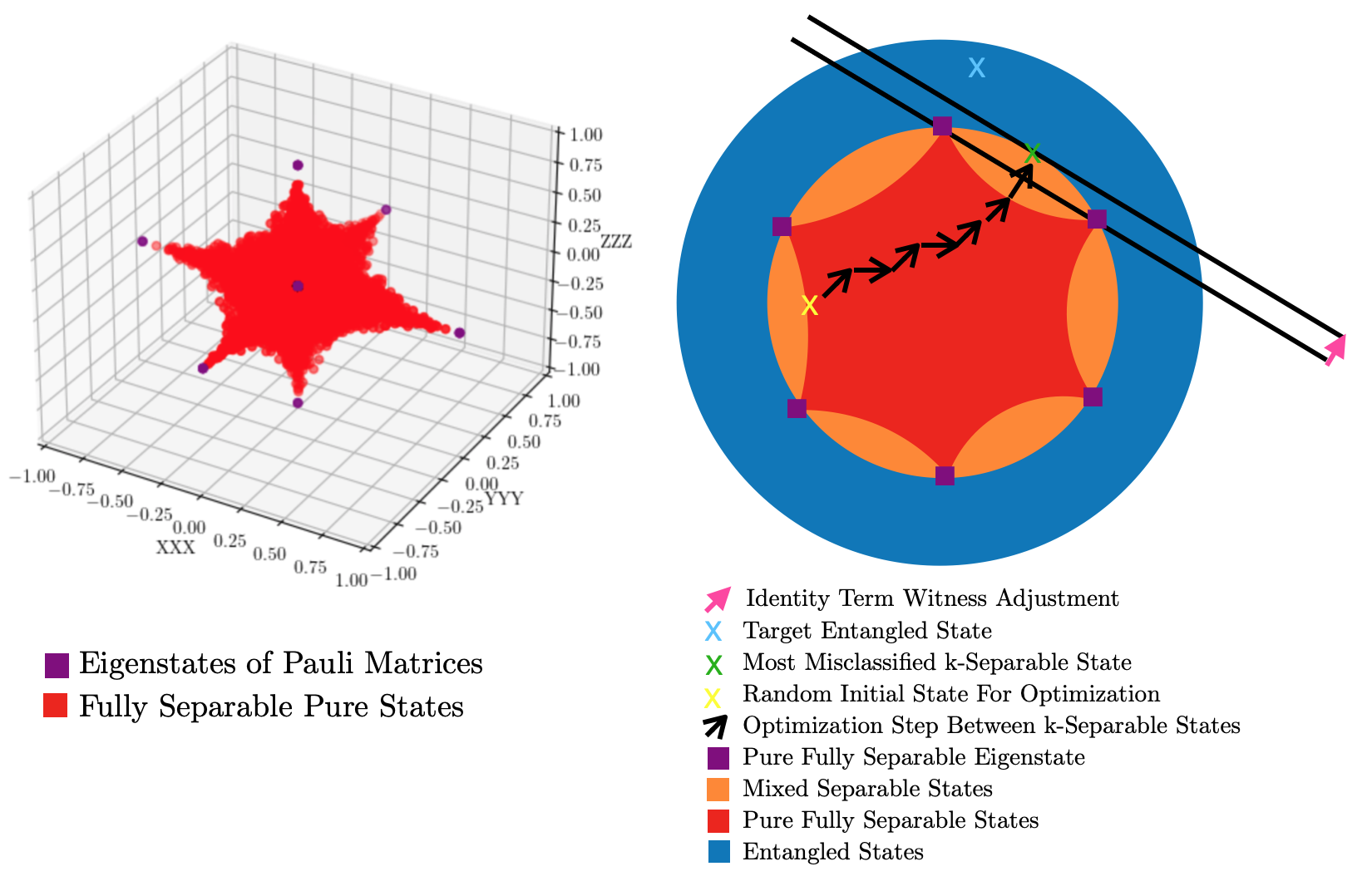}
\caption{Left: In red: A sampling of \(10^5\) fully separable 3 qubit pure states plotted against their expectation values for Pauli operators XXX, YYY, and ZZZ. In purple: the fully separable pure eigenstates are visible at the extreme points.
Right: Conceptual geometry of of feature space, including k-separable states and entangled states. Steps of differential mixed state optimization, and the modification of the bias term in the witness are also shown. The head and tail of each optimization step arrow would be the position of a k-separable state in feature space which is then adjusted to move toward the witness. Note the plane before adjustment shows, for clarity, the worst case scenario where after training with the eigenstates, the plane touches two of them. Typically there would initially be a small margin. Also note that the plane before adjustment is aligned such that the optimizer does not converge to the separable state closest to the target, which is typically the case.}
\label{fig: e_o}
\end{figure*}

Given that we have chosen feature space to be defined by the expectation values of generalized Pauli operators, it is easy to show that the eigenstates of these operators must constitute extreme points in that space.
Namely, any separable states with corresponding feature vectors \(f\) which have any elements \(f_i = \pm 1\), which are the maximum and minimum values of \(\langle \psi | \sigma_k | \psi \rangle\)
for any generalized Pauli Matrix
\(\sigma_k\) must be extreme points of the separable states in feature space. 
We consider only pure states \(|\psi_a\rangle\). Since the result of
\(\sigma_k |\psi _a\rangle\)
is some other pure state
\( \psi_b\),
we only want to optimize the inner product
\(\langle \psi_a | \psi_b \rangle\). As Pauli matrices preserve norms, and have eigenvalues of \(\pm 1\), we can just choose $\vert\psi_a\rangle$ as their eigenstates. 
Plotting these eigenstates next to a sampling of the pure
fully separable states revealed that they do form visible extreme points, as shown in Figure \ref{fig: e_o}.

Our method requires less tuning of hyperparameters when we sample five additional fully separable training states near each eigenstate. It is possible to train SVMs to produce similar witnesses with the eigenstates as the only separable training data, but this requires significant adjustment of the learning rate, batch size, and regularization terms far from their default settings provided by Tensorflow \cite{TensorFlow}. Comparing witnesses trained with a GHZ target state and these separable states to known stabilizer witnesses, we found that only the bias term needed an adjustment, indicating that the witness was oriented almost correctly in feature space but not translated properly. The slight differences in orientation are listed in section \ref{sec:conf}. We believe that the pure separable eigenstates offer a symmetric set of extreme points which force the SVM training program to identify the orientation of their surface, but are possibly recessed somewhat from the immediate surface of feature vectors for separable states near the target. Conversely, while the performance would slightly improve sampling more than 5 additional points, we decided to limit the burden of additional training data. By choosing how many additional points to sample and how they are dispersed, there are many available ways to adjust this method to a given scenario depending on state size, computational resources, and the particular target state.

The 5 additional training states per Pauli eigenstate that we include in our training set are found through application of a randomly sampled but narrowly distributed set of unitary matrices. For each qubit we apply an operator $H_{i}$ so that the total $n$ qubit system operator is given by $H = \bigotimes_i H_i$ where

\begin{equation} \label{eq:haar}  H_i = \begin{pmatrix} e^{-i\sigma\frac{(\phi_i + \omega_i)}{2}}\cos(\frac{\sigma\theta_i}{2}) && -e^{i\sigma\frac{(\phi_i - \omega_i)}/{2}}\sin(\frac{\sigma\theta_i}{2}) \\e^{-i\sigma\frac{\phi_i -\omega_i}{2}}\sin(\frac{\sigma\theta_i}{2}) && e^{i\sigma\frac{(\phi_i + \omega_i)}{2}}\cos(\frac{\sigma\theta_i}{2})
\end{pmatrix}. \end{equation}
Here \(\theta_i \in [0, \pi] \) are identically independently distributed according to the probability density function \(\frac{1}{2} \sin \theta_i\), \(\phi_i \) and \(\omega_i \) follow independent uniform distributions in \([0,2\pi]\) \cite{haar}, and we choose \(\sigma\) to be fixed at \(0.05\). Our results are relatively insensitive to small changes to \(\sigma\), and larger changes to \(\sigma\) can likely be corrected via hyperparameter tuning.
We chose the value of \(\sigma\) through trial and error by comparing GHZ witnesses we generate with those derived through the stabilizer formalism \footnote[1]{We consider our approach to have a better performance when the ratios between the bias term and every other term is closer to the same ratio for those coefficients from the formalism, as all coefficients can be scaled up or down together without defining an effectively different hyperplane in feature space. For example, the 3-qubit GHZ witness can be written as \(4III - XXX + XYY + YXY + YYX\)\, but a hypothetical witness derived with our method might be written as \(2III - 0.5XXX + 0.5XYY + 0.5YXY + 0.51YYX\). We expect the ratio of the bias term to all other terms to be 4, but in this example the hypothetical witness has a ratio of 3.92 between the \(YYX\) and \(III\) terms. We would quote the error as 1-3.92/4 \(\approx\) 0.02. Our method of measuring error could be optimized more precisely in future work.}.

Overall, we use
\(10^3\)
entangled training states and
\(6^{n+1}\)
separable training states for a system of
\(n\)
qubits. One additional state is used to correct the bias term through differential mixed state optimization, discussed in section \ref{sec:mso} below.
Comparing to the original SVM approach \cite{PhysRevApplied.19.034058}, where
\(5 \times 10^5\)
separable training states were used, if the same density of points in feature
space were to be maintained, they would have to fill a cube of side length 2 in a \(4^N - 1\) dimensional space, as there are \(4^N\) distinct generalized Pauli operators, but the identity operator has an expectation value of 1 for all states, while all others can vary independently. Then we can say that the size of training data should scale roughly as \(5 \times 10^5 \times 2^{4^N - 4^3}\). In the case of 4 qubits, the original method would then already require on the order of \(10^{63}\) training states, whereas this new method needs only 7776, representing a reduction of almost sixty orders of magnitude. In the case of 5 qubits, the original method would need more than \(10^{294}\), whereas we trained 5 qubit witnesses with 46657 states.

Comparing to a more recent SVM paper \cite{new} we find that our
method still requires less than 40\% of their training data, and is conceptually much simpler to construct. They provide no argument for how their training data scales with \(N\). With our method, no new constructions are required to extend our work to larger systems (as we demonstrate with 5 qubits) whereas that method relies on separate work \cite{new_4qubit_ref} specific to the space of 4 qubits.

\section{\label{sec:mso} Differential Mixed State Optimization}

Building on the intuition of Fig. \ref{fig: e_o} where the Pauli eigenstates help orient our witness plane in this section, we describe a method based on differential mixed state optimization for biasing this plane such that it does not misclassify separable states as entangled. In particular, we find a separable state with the lowest expectation value possible for a hyperplane we trained using an SVM and the training data described in section \ref{sec:newdata}. We then correct the witness by subtracting this expectation value from the bias term. This guarantees all separable states have a nonnegative expectation value while also ensuring our final mixed state from the optimization has an expectation value of 0, which maximizes the noise tolerance.

Note that pure separable states for N qubits can in general be specified as k-separable states, where k refers to the number of groups of qubits not entangled with any others. This restricts k to the range \([1,N]\), where for instance a 1-separable state suggests that N-1 qubits are entangled with each other while 1 is not, whereas an N-separable state is considered fully separable in the sense that none of the qubits are entangled. A mixed k-separable state is then a convex combination of density matrices representing k-separable pure states.

Given such a mixed state which we parameterize by a certain set of parameters, we consider the gradient of the expectation value of the witness with respect to those parameters and decide whether to increase or decrease each by a small step size, thereby slightly changing the state. Our parameterization of mixed states is described in Appendix \ref{sec:mso_parameterization}. This makes our implementation an example of a differential program \cite{diff}, distinct from other optimization frameworks such as semidefinite programming. Figure \ref{fig: e_o} from section III illustrates the procedure this program follows.

We used Tensorflow to implement gradient descent across k-separable states. Gradient descent and our implementation are briefly described in Appendix \ref{sec:gradient_descent}. Figure \ref{fig: optimization run} shows the loss for an optimization  of a mixed state in terms of expectation value of a GHZ state witness, plotted against the number of iterations of the optimizer. The orange line shows the lowest expectation value among the fully separable eigenstate of the generalized Pauli matrices which we use for training. The loss decreases continuously to an apparent minimum in less than 120 iterations. For 5 qubits, this process always takes less than 5 minutes on the freely available version of Google Colaboratory for a random initial state.

\begin{figure*}
\includegraphics[width=4.5 in,height=3.5 in]{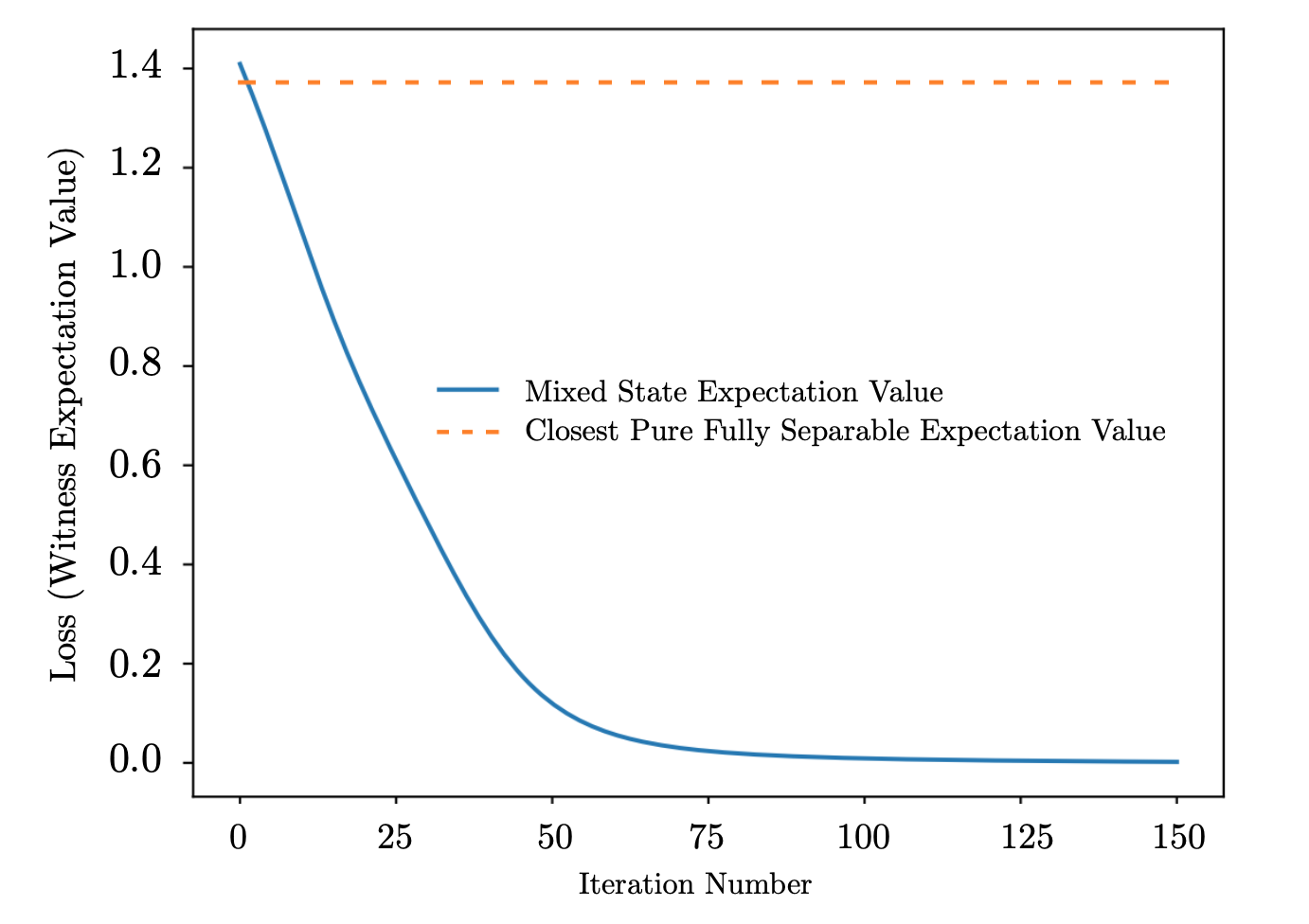}
\caption{Results from one optimization run. Compare the expectation value of the k-separable mixed state being optimized to the expectation value of the closest pure fully separable eigenstate. We can find k-separable states with a much lower expectation value, and we use the final expectation value to modify the bias term of the witness.}
\label{fig: optimization run}
\end{figure*}

To implement the optimizer we parameterized sets of pure k-separable states for 4 and 5 qubits. We then combined the density matrices for these pure states in convex combinations using additional parameters to create mixed k-separable states. This parameterization is defined in Appendix \ref{sec:mso_parameterization}. For each step the optimizer took, we constructed such a state, and computed its expectation value according to a witness generated using the method described in section \ref{sec:newdata}

Using the eigenstates of the generalized Pauli matrices, as well as \(10^3\) entangled Werner
training states described with Equation \ref{eq: noise}, we can produce GHZ witnesses
for three, four, and five qubits for which all terms but the bias term match those of the stabilizer formalism in terms of their ratios to each other. To fix the identity
term, the coefficients and the labels of the nonzero features are passed
to a mixed state optimizer, which then searches through k-separable
mixed states with the above parameterization using the expectation value
of the witness as the loss function.

Recall that all separable states should have positive expectation
values, but to maximize the noise tolerance, we would want the decision
boundary to just touch the surface of the separable states. Since our
parameterization seems to span the entire space of separable states, and
the space of separable mixed states is convex, we believe our optimizer
can find the most misclassified separable state up to arbitrary
precision by minimizing the expectation value of our witness.
Once this state is found, its expectation value is subtracted from the
bias term, which brings the decision boundary right to the surface
of the separable states. Figure \ref{fig: e_o} in section \ref{sec:newdata} illustrates the correction in a cartoon.

As described in Appendix \ref{sec:mso_parameterization} we use a subset of all permutations of all partitions of pure states in the size of the system we are considering. The number of partitions \(p\) for natural numbers \(n\) scales as \cite{partitions}
\begin{equation} \label{eq: partitions_formula}
p(n) \sim \frac{1}{4n\sqrt{3}}\exp{\left[\pi \sqrt{\frac{2n}{3}}\right]}
\end{equation}
and we consider less than the total possible \(n!\) permutations of each partition by ignoring permutations within groups of entangled qubits. Finally, as we discuss in Appendix \ref{sec:mso_parameterization}, for \(N\) qubits we consider less than \(2^N\) copies of each permutation of each partition, so as to guarantee we can reach the maximally mixed state for any subset of fewer than \(N\) qubits. For an \(N\) qubit state we represent it with \(2^{N+1}\) 64-bit floating point numbers. So for a \(N\) qubit optimizer, the memory load \(M(N)\), in bytes, which must be stored in RAM should be bounded by

\begin{equation} \label{eq: memory_requirement}
M(N) \leq \frac{2^{2N+2} (N-1)!}{\sqrt{3}}\exp{\left[\pi \sqrt{\frac{2N}{3}}\right]}
\end{equation}

With 64 GB of available random access memory storage, it should be possible to create a functioning optimizer using the technique described here for systems of 6 and 7 qubits. 

\section{\label{sec:conf} Confirmation of 4 and 5 Qubit GHZ Witnesses}

Witnesses for GHZ states in systems of arbitrarily many qubits can be
derived from the stabilizer formalism \cite{toth} based on
the expectation value of the Bell operator for
\( N \)
qubits defined as
\begin{equation} \label{eq: stabilizer} 
\begin{split}
M_N  := \frac{1}{2^{N-1}} (X^{(1)}X^{(2)}X^{(3)}X^{(4)}...X^{(N-1)}X^{(N)}\\-Y^{(1)}Y^{(2)}X^{(3)}X^{(4)}...
X^{(N-1)}X^{(N)}\\+Y^{(1)}Y^{(2)}Y^{(3)}Y^{(4)}...X^{(N-1)}X^{(N)})
\end{split}
\end{equation} 

where \(X^{(i)}\) refers to the operator \(X\) acting on the \(i\)th qubit,
\(X^{(i)}Y^{(j)}\)refers to \(X^{(i)} \otimes Y^{(j)}\),
and each term in  refers to the sum of all its
possible permutations.

We have \(\langle M_N \rangle \leq \frac{1}{2}\) for k-separable states \cite{toth} which can be rewritten as

\begin{widetext}
\begin{equation} \label{eq: W_mermin} 
\begin{split}
     \langle \mathcal{W}_{Mermin}^{N} \rangle = \langle 2^{N-2}I^{\otimes N} -X^{(1)}X^{(2)}X^{(3)}X^{(4)}...X^{(N-1)}X^{(N)}\\+Y^{(1)}Y^{(2)}X^{(3)}X^{(4)}... X^{(N-1)}X^{(N)}\\-Y^{(1)}Y^{(2)}Y^{(3)}Y^{(4)}...X^{(N-1)}X^{(N)} \rangle \geq 0
\end{split}
\end{equation} 
\end{widetext}

where
(\ref{eq: W_mermin}) is the N-qubit GHZ Mermin witness. Importantly the inequality in (\ref{eq: W_mermin}) still holds when both sides of the equation are multiplied by a constant
factor, meaning that only the ratio in magnitudes between the
\(I^{\otimes N}\)
terms and every other term as well as the signs of each term in the sum
are relevant.

Using our new method, we trained SVMs with the Mermin witness feature subset from (\ref{eq: W_mermin}) for systems of 4 and 5 qubits with the GHZ state as the target. The coefficients we found and the coefficients found through the stabilizer formalism are listed below in Tables \ref{tab: 4 qubit GHZ witness} and \ref{tab: 5 qubit GHZ witness}. For ease of comparison, the coefficients have been normalized such that the SVM-derived identity coefficients match the identity coefficients from the stabilizer formalism.

For 4 qubits, the largest error for any coefficient was about 6.5\%, while for 5 qubits, the largest error was 0.9\%. We don't believe there is necessarily a trend toward lower error for larger systems, as the error is due to differences in the terms besides the bias term which vary based on hyperparameter tuning during SVM training or our choices of separable training data. 

\begin{table}[b]
\caption{\label{tab: 4 qubit GHZ witness} 4 Qubit SVM Derived GHZ Witness Coefficients and Stabilizer Formalism Coefficients}
\begin{ruledtabular}
\begin{tabular}{c c c c}
\textbf{Feature} & \textbf{Mermin}  &  \textbf{SVM Derived} & \textbf{Percent} \\
& \textbf{Witness} & \textbf{Coefficient} & \textbf{Error} \\
& \textbf{Coefficient} &  & \\
\hline
\textbf{IIII} & 4 & 4 & Normalized to 0 \\
\textbf{XXXX} & -1 & -0.935 & -6.46 \\
\textbf{XXYY} & 1 & 1.024 & 2.40 \\
\textbf{XYXY} & 1 & 1.023 & 2.30 \\
\textbf{XYYX} & 1 & 1.024 & 2.36 \\
\textbf{YXXY} & 1 & 1.023 & 2.29 \\
\textbf{YXYX} & 1 & 1.024 & 2.40 \\
\textbf{YYXX} & 1 & 1.021 & 2.07 \\
\end{tabular}
\end{ruledtabular}
\end{table}

\begin{table}[b]
\caption{\label{tab: 5 qubit GHZ witness} 4 Qubit SVM Derived GHZ Witness Coefficients and Stabilizer Formalism Coefficients}
\begin{ruledtabular}
\begin{tabular}{cccc}
\textbf{Feature} & \textbf{Mermin}  &  \textbf{SVM Derived} & \textbf{Percent} \\
& \textbf{Witness} & \textbf{Coefficient} & \textbf{Error} \\
& \textbf{Coefficient} &  & \\
\hline
\textbf{IIIII} & 8 & 8 & Normalized to 0 \\
\textbf{XXXXX} & -1 & -1.0013 & 0.13 \\
\textbf{YYXXX} & 1 & 1.0023 & 0.23 \\
\textbf{YXYXX} & 1 & 1.0005 & 0.05 \\
\textbf{YXXYX} & 1 & 1.0038 & 0.38 \\
\textbf{YXXXY} & 1 & 1.0036& 0.36 \\
\textbf{XYYXX} & 1 & 0.9999 & -0.01\\
\textbf{XYXYX} & 1 & 1.0019 & 0.19\\
\textbf{XYXXY} & 1 & 1.0062 & 0.62\\
\textbf{XXYYX} & 1 & 1.0043 & 0.43 \\
\textbf{XXYXY} & 1 & 1.0095 & 0.95 \\
\textbf{XXXYY} & 1 & 1.0008 & 0.08\\
\textbf{XYYYY} & -1 & -1.0070 & 0.70 \\
\textbf{YXYYY} & -1 & -0.9990 & -0.10 \\
\textbf{YYXYY} & -1 & -0.9986& -0.14 \\
\textbf{YYYXY} & -1 & -0.9948 & -0.52\\
\textbf{YYYYX} & -1 & -0.9980 & -0.20 \\
\end{tabular}
\end{ruledtabular}
\end{table}

\section{\label{sec:rfe} Recursive Feature Elimination for 4 and 5 Qubit W State Witnesses}

We have demonstrated that our approach recovers approximately the correct ratio between coefficients of entanglement witnesses found using stabilizer formalism. We will now extend this work to the derivation of novel witnesses through recursive feature elimination (RFE) to identify and remove the terms of our witnesses that least impact the outcome. In particular, we ignore terms with coefficients below a cutoff, which is chosen through trial and error to preserve the noise tolerance of the initial witness.

To derive new W state witnesses, RFE was used to train witnesses with the largest noise tolerance possible for smaller sets of features. Suppose after this step there are \(T\) nonzero features. We start by choosing the best feature to eliminate. Witnesses are trained with all possible subsets of \(T-1\) features with the exception of the subset excluding the bias term, as we always need to adjust the bias term to ensure witnesses are valid.  The noise tolerance of each witness is then computed by finding the largest value of \(p\) in (\ref{eq: noise}) for which the witness has a negative expectation value, as described in section \ref{sec:old_scheme_SVMs}. The feature subset corresponding as the witness with the highest noise tolerance is selected. Then every subset with \(T-2\) of this feature subset is considered, new witnesses are trained, and the procedure is repeated until we find the witness with the largest noise tolerance of some target number of features. This method can also be run for much longer to find the overall largest noise tolerance of any witness for a particular target by continuing to remove features until only the bias term remains. 

In general, as adjusting the hyperparameters for the SVM learning algorithm for every witness and every set of features would be challenging, it is more effective to first remove features and see if the noise tolerance can be improved rather than try to find the best result for a given number of features. This makes RFE an efficient method not just for reducing the number of measurements, but also for raising the noise tolerance.

We find that for 4 qubits, training a witness with feature subsets produces higher noise tolerances than training without restriction. Presented in Appendix \ref{sec:coeff} are tables \ref{tab: W witness, 46 terms}, \ref{tab: W witness, 38 terms}, \ref{tab: W witness 28 terms}, \ref{tab: W witness, 20 terms} and \ref{tab: W witness, 180 terms} which show 4 and 5 qubit W state witnesses found using RFE. For 4 qubits, before any features were removed, the W state witness had 46 terms; for 5 qubits it had 180 terms. 

By training with the full set of features and tuning the hyperparameters, we were able to achieve similar margins between our entangled training data and the Pauli eigenstates as we found in our 4 qubit witness with the highest final noise tolerance. However, we could not achieve the same final noise tolerance after the bias term adjustment. Furthermore, while we could tune the hyperparameters to give us a witness with the same number of non-negligible features, different features were always present. We believe that only by removing features will the optimal hyperplane found by the SVM be identical to the optimal witness after the bias term is adjusted. There might exist several similar or near-optimal hyperplanes which the SVM does not converge to if it can consider all features.

Note that in the case of 4 qubit W state witnesses generated with our method, coefficients smaller than a cutoff of 0.02 could be ignored without changing the noise tolerance. But in the case of 5 qubit W state witnesses, we had to include terms as small as 0.0004. Note both of these values apply to coefficients after the all terms have been normalized such that the bias term coefficient is exactly 1. This cutoff value was found with trial and error and may need to be further optimized.

W state fidelity witnesses in systems of \(N\) qubits have been shown \cite{guhne} to have noise tolerances of
\begin{equation} \label{eq: W_fidelity_noise_tolerance} p_{max} = \frac{1}{N}\times \frac{1}{1-\frac{1}{2^N}}.\end{equation}
However, this noise tolerance is not optimal \cite{PhysRevApplied.19.034058}\cite{taming}. The 4 qubit fidelity witness has a noise tolerance of \(\frac{4}{15}\) or 0.27. Our 4 qubit witnesses have noise tolerances of about 0.30,
0.40, and 0.46, for 46, 38 and 28 terms respectively. Curiously, the noise tolerance increases, and this seems to correspond with the feature \(ZZZZ\) and permutations of \(IZZZ\) being removed. Note that while each of these witnesses has a larger noise tolerance than the witness given by the fidelity method, but smaller than the limit of 0.526 shown by \cite{taming}. 

With 5 qubits, an analytical method of witness construction known as the fidelity method \cite{fidelity} results in a witness with a noise tolerance of about 0.21.  Our method method produces a witness with 180 terms and a noise tolerance of 0.15 and another 20 terms and a noise tolerance at 0.05.

To numerically find the noise tolerance after training is completed, we first note the target \(p_e\) which the witness is trained to detect. In step sizes of 0.001, we vary \(p\) in (\ref{eq: noise}) to compute new density matrices \(\rho\). We compute the expectation values of our witness \(\mathcal{W}\) at each step, stopping when they become positive. The last value of \(p\) for which \(Tr(\rho \mathcal{W})\) is positive is reported as the noise tolerance.

\section{\label{sec:verif} Verification of Generated W State Witnesses}
\subsection{\label{sec:numerical_verification} Numerical Test Data Generation and Testing}
To generate entangled testing data, we produced
\(10^6\)
states using Equation \ref{eq: noise} with
\(p\)
varying between 0 and 0.25 or 0 and the noise tolerance quoted according to (\ref{eq: entangled_label}) in section \ref{sec:rfe}.

We used a method similar to that of \cite{PhysRevApplied.19.034058} to
produce separable and k-separable testing data. As noted in Appendix \ref{sec:mso_parameterization}, any four qubit pure separable state can be constructed as the
tensor product of either an arbitrary single qubit and three qubit pure state or two independent two qubit pure states. Note the case of more than one separable qubit is also included by considering arbitrary two or three qubit subsystems. For example, a pure fully separable four qubit state could be represented as a one and three qubit subsystem, where the three qubit subsystem happens to be in a fully separable pure state itself.
These separable states
can then be permuted according to all the arrangements in Appendix \ref{sec:perms}.
Similarly any five qubit mixed state can be generated with combinations
of arbitrary 1,2,3 and 4 qubit states and permuted according to Appendix C.

Each mixed state used for testing was generated according to an MA
distribution, in which the density matrices for \(2^{N}\)
pure \(N\) qubit states were combined in a convex combination with coefficients composing a vector \textbf{x} defined by a symmetric Dirichlet distribution \cite{PhysRevApplied.19.034058} as defined as 
\begin{equation} \label{eq: dirichlet} 
\text{Dir} (x| \alpha) = \frac{\Gamma({N\alpha})}{\left(\Gamma(\alpha)\right)^N} \prod_{j=1}^{N}x_j^{\alpha-1}
\end{equation}
While the choice of \(\alpha\) is arbitrary, smaller values of
\(\alpha\) have been shown to lead to states with higher purity on average, which was in turn found in the case of 3 qubits to correspond to states with higher fidelity to the target W state and so provide the best chance for misclassification by the witness \cite{PhysRevApplied.19.034058}.

We generated \(10^6\) of these separable states for 4 and 5 qubits. Due to limited computational resources, we could not test with much larger sample sizes, so additional verification may be necessary to confirm the validity of these witnesses.  As well, the choice of \(\alpha\)
is arbitrary, and perhaps more stringent testing data could be generated with some other value or a different distribution entirely.

We must also note that physical verification of these witnesses should be completed as well, in a similar manner to that which was carried out for the original 3 qubit SVM derived witnesses \cite{PhysRevApplied.19.034058}. This verification would require new quantum circuits for constructing test states and implementing the witness, and was deemed out of the scope of this project. 

Granted these caveats, we propose the witnesses in Appendix \ref{sec:coeff} as valid W state witnesses for 4 and 5 qubits, as they correctly classify 100\% of our testing data.

\section{\label{sec:conclusion} Conclusion}

In this work we first introduced a simple method for generating small training data sets for SVMs which can derive an approximate entanglement witness. We then introduced a method, named Mixed State Optimization, for modifying the bias terms of the approximate witness to produce a witness that correctly classifies \textit {all} separable states while maximizing noise tolerance.

We demonstrate that our method scales dramatically better than the original SVM method \cite{PhysRevApplied.19.034058}, due to the use of a subset of extreme points corresponding to the eigenstates of the Pauli operators as the original training states. We achieved an exponential reduction of training data size. We then validate this conceptually simple and scaleable approach by reproducing the coefficients of 4 and 5 qubit GHZ witnesses derived from the stabilizer formalism, given the same feature subset. The largest deviation from the theoretical witness for any coefficient of 4-qubit witness we found was about 6.5\%, while all deviations from the 5-qubit GHZ witness were less than 1\%.

We also produced a set of three novel 4-qubit W-state witnesses, and two 5-qubit W-state witnesses. We then verified these witnesses using testing data numerically generated from an MA distribution which allows for the random generation of states near the witness boundary. We also proposed computational and physical methods for further verification of our results. 

We expect this method to be useful for generating entanglement witnesses for a variety of target states and for larger systems. As the training data sizes scale more slowly than with existing methods, and mixed state optimizers written in Tensorflow can easily handle many thousands or millions of parameters \cite{adam}, this method is directly applicable to higher dimensional systems.

\bibliography{b}

\begin{thebibliography}{35}%
\makeatletter
\providecommand \@ifxundefined [1]{%
 \@ifx{#1\undefined}
}%
\providecommand \@ifnum [1]{%
 \ifnum #1\expandafter \@firstoftwo
 \else \expandafter \@secondoftwo
 \fi
}%
\providecommand \@ifx [1]{%
 \ifx #1\expandafter \@firstoftwo
 \else \expandafter \@secondoftwo
 \fi
}%
\providecommand \natexlab [1]{#1}%
\providecommand \enquote  [1]{``#1''}%
\providecommand \bibnamefont  [1]{#1}%
\providecommand \bibfnamefont [1]{#1}%
\providecommand \citenamefont [1]{#1}%
\providecommand \href@noop [0]{\@secondoftwo}%
\providecommand \href [0]{\begingroup \@sanitize@url \@href}%
\providecommand \@href[1]{\@@startlink{#1}\@@href}%
\providecommand \@@href[1]{\endgroup#1\@@endlink}%
\providecommand \@sanitize@url [0]{\catcode `\\12\catcode `\$12\catcode `\&12\catcode `\#12\catcode `\^12\catcode `\_12\catcode `\%12\relax}%
\providecommand \@@startlink[1]{}%
\providecommand \@@endlink[0]{}%
\providecommand \url  [0]{\begingroup\@sanitize@url \@url }%
\providecommand \@url [1]{\endgroup\@href {#1}{\urlprefix }}%
\providecommand \urlprefix  [0]{URL }%
\providecommand \Eprint [0]{\href }%
\providecommand \doibase [0]{https://doi.org/}%
\providecommand \selectlanguage [0]{\@gobble}%
\providecommand \bibinfo  [0]{\@secondoftwo}%
\providecommand \bibfield  [0]{\@secondoftwo}%
\providecommand \translation [1]{[#1]}%
\providecommand \BibitemOpen [0]{}%
\providecommand \bibitemStop [0]{}%
\providecommand \bibitemNoStop [0]{.\EOS\space}%
\providecommand \EOS [0]{\spacefactor3000\relax}%
\providecommand \BibitemShut  [1]{\csname bibitem#1\endcsname}%
\let\auto@bib@innerbib\@empty
%</preamble>
\bibitem [{\citenamefont {Ali-Khan}\ \emph {et~al.}(2007)\citenamefont {Ali-Khan}, \citenamefont {Broadbent},\ and\ \citenamefont {Howell}}]{PhysRevLett.98.060503}%
  \BibitemOpen
  \bibfield  {author} {\bibinfo {author} {\bibfnamefont {I.}~\bibnamefont {Ali-Khan}}, \bibinfo {author} {\bibfnamefont {C.~J.}\ \bibnamefont {Broadbent}},\ and\ \bibinfo {author} {\bibfnamefont {J.~C.}\ \bibnamefont {Howell}},\ }\bibfield  {title} {\bibinfo {title} {Large-alphabet quantum key distribution using energy-time entangled bipartite states},\ }\href {https://doi.org/10.1103/PhysRevLett.98.060503} {\bibfield  {journal} {\bibinfo  {journal} {Phys. Rev. Lett.}\ }\textbf {\bibinfo {volume} {98}},\ \bibinfo {pages} {060503} (\bibinfo {year} {2007})}\BibitemShut {NoStop}%
\bibitem [{\citenamefont {Zhong}\ \emph {et~al.}(2015)\citenamefont {Zhong}, \citenamefont {Zhou}, \citenamefont {Horansky}, \citenamefont {Lee}, \citenamefont {Verma}, \citenamefont {Lita}, \citenamefont {Restelli}, \citenamefont {Bienfang}, \citenamefont {Mirin}, \citenamefont {Gerrits},\ and\ \citenamefont {et~al.}}]{zhong}%
  \BibitemOpen
  \bibfield  {author} {\bibinfo {author} {\bibfnamefont {T.}~\bibnamefont {Zhong}}, \bibinfo {author} {\bibfnamefont {H.}~\bibnamefont {Zhou}}, \bibinfo {author} {\bibfnamefont {R.~D.}\ \bibnamefont {Horansky}}, \bibinfo {author} {\bibfnamefont {C.}~\bibnamefont {Lee}}, \bibinfo {author} {\bibfnamefont {V.~B.}\ \bibnamefont {Verma}}, \bibinfo {author} {\bibfnamefont {A.~E.}\ \bibnamefont {Lita}}, \bibinfo {author} {\bibfnamefont {A.}~\bibnamefont {Restelli}}, \bibinfo {author} {\bibfnamefont {J.~C.}\ \bibnamefont {Bienfang}}, \bibinfo {author} {\bibfnamefont {R.~P.}\ \bibnamefont {Mirin}}, \bibinfo {author} {\bibfnamefont {T.}~\bibnamefont {Gerrits}},\ and\ \bibinfo {author} {\bibnamefont {et~al.}},\ }\bibfield  {title} {\bibinfo {title} {Photon-efficient quantum key distribution using time–energy entanglement with high-dimensional encoding},\ }\href {https://doi.org/10.1088/1367-2630/17/2/022002} {\bibfield  {journal} {\bibinfo  {journal} {New Journal of Physics}\ }\textbf {\bibinfo {volume} {17}},\
  \bibinfo {pages} {022002} (\bibinfo {year} {2015})}\BibitemShut {NoStop}%
\bibitem [{\citenamefont {Chen}\ \emph {et~al.}(2014)\citenamefont {Chen}, \citenamefont {Lei},\ and\ \citenamefont {Romero}}]{chen}%
  \BibitemOpen
  \bibfield  {author} {\bibinfo {author} {\bibfnamefont {L.}~\bibnamefont {Chen}}, \bibinfo {author} {\bibfnamefont {J.}~\bibnamefont {Lei}},\ and\ \bibinfo {author} {\bibfnamefont {J.}~\bibnamefont {Romero}},\ }\bibfield  {title} {\bibinfo {title} {Quantum digital spiral imaging},\ }\bibfield  {journal} {\bibinfo  {journal} {Light: Science Applications}\ }\textbf {\bibinfo {volume} {3}},\ \href {https://doi.org/10.1038/lsa.2014.34} {10.1038/lsa.2014.34} (\bibinfo {year} {2014})\BibitemShut {NoStop}%
\bibitem [{\citenamefont {Lukens}\ \emph {et~al.}(2020{\natexlab{a}})\citenamefont {Lukens}, \citenamefont {Lu}, \citenamefont {Qi}, \citenamefont {Lougovski}, \citenamefont {Weiner},\ and\ \citenamefont {Williams}}]{lukens}%
  \BibitemOpen
  \bibfield  {author} {\bibinfo {author} {\bibfnamefont {J.~M.}\ \bibnamefont {Lukens}}, \bibinfo {author} {\bibfnamefont {H.-H.}\ \bibnamefont {Lu}}, \bibinfo {author} {\bibfnamefont {B.}~\bibnamefont {Qi}}, \bibinfo {author} {\bibfnamefont {P.}~\bibnamefont {Lougovski}}, \bibinfo {author} {\bibfnamefont {A.~M.}\ \bibnamefont {Weiner}},\ and\ \bibinfo {author} {\bibfnamefont {B.~P.}\ \bibnamefont {Williams}},\ }\bibfield  {title} {\bibinfo {title} {All-optical frequency hopping and broadcasting in wavelength-multiplexed channels},\ }in\ \href {https://doi.org/10.1364/CLEO_SI.2020.SF2L.2} {\emph {\bibinfo {booktitle} {Conference on Lasers and Electro-Optics}}}\ (\bibinfo  {publisher} {Optica Publishing Group},\ \bibinfo {year} {2020})\ p.\ \bibinfo {pages} {SF2L.2}\BibitemShut {NoStop}%
\bibitem [{\citenamefont {Thomas}\ \emph {et~al.}(2022)\citenamefont {Thomas}, \citenamefont {Ruscio}, \citenamefont {Morin},\ and\ \citenamefont {Rempe}}]{thomas}%
  \BibitemOpen
  \bibfield  {author} {\bibinfo {author} {\bibfnamefont {P.}~\bibnamefont {Thomas}}, \bibinfo {author} {\bibfnamefont {L.}~\bibnamefont {Ruscio}}, \bibinfo {author} {\bibfnamefont {O.}~\bibnamefont {Morin}},\ and\ \bibinfo {author} {\bibfnamefont {G.}~\bibnamefont {Rempe}},\ }\bibfield  {title} {\bibinfo {title} {Efficient generation of entangled multiphoton graph states from a single atom},\ }\href {https://api.semanticscholar.org/CorpusID:249062813} {\bibfield  {journal} {\bibinfo  {journal} {Nature}\ }\textbf {\bibinfo {volume} {608}},\ \bibinfo {pages} {677 } (\bibinfo {year} {2022})}\BibitemShut {NoStop}%
\bibitem [{\citenamefont {Kues}\ \emph {et~al.}(2017)\citenamefont {Kues}, \citenamefont {Reimer}, \citenamefont {Roztocki}, \citenamefont {Cortés}, \citenamefont {Sciara}, \citenamefont {Wetzel}, \citenamefont {Zhang}, \citenamefont {Cino}, \citenamefont {Chu}, \citenamefont {Little},\ and\ \citenamefont {et~al.}}]{kues}%
  \BibitemOpen
  \bibfield  {author} {\bibinfo {author} {\bibfnamefont {M.}~\bibnamefont {Kues}}, \bibinfo {author} {\bibfnamefont {C.}~\bibnamefont {Reimer}}, \bibinfo {author} {\bibfnamefont {P.}~\bibnamefont {Roztocki}}, \bibinfo {author} {\bibfnamefont {L.~R.}\ \bibnamefont {Cortés}}, \bibinfo {author} {\bibfnamefont {S.}~\bibnamefont {Sciara}}, \bibinfo {author} {\bibfnamefont {B.}~\bibnamefont {Wetzel}}, \bibinfo {author} {\bibfnamefont {Y.}~\bibnamefont {Zhang}}, \bibinfo {author} {\bibfnamefont {A.}~\bibnamefont {Cino}}, \bibinfo {author} {\bibfnamefont {S.~T.}\ \bibnamefont {Chu}}, \bibinfo {author} {\bibfnamefont {B.~E.}\ \bibnamefont {Little}},\ and\ \bibinfo {author} {\bibnamefont {et~al.}},\ }\bibfield  {title} {\bibinfo {title} {On-chip generation of high-dimensional entangled quantum states and their coherent control},\ }\href {https://doi.org/10.1038/nature22986} {\bibfield  {journal} {\bibinfo  {journal} {Nature}\ }\textbf {\bibinfo {volume} {546}},\ \bibinfo {pages} {622–626} (\bibinfo {year}
  {2017})}\BibitemShut {NoStop}%
\bibitem [{\citenamefont {Imany}\ \emph {et~al.}(2018)\citenamefont {Imany}, \citenamefont {Odele}, \citenamefont {Jaramillo-Villegas}, \citenamefont {Leaird},\ and\ \citenamefont {Weiner}}]{imany}%
  \BibitemOpen
  \bibfield  {author} {\bibinfo {author} {\bibfnamefont {P.}~\bibnamefont {Imany}}, \bibinfo {author} {\bibfnamefont {O.~D.}\ \bibnamefont {Odele}}, \bibinfo {author} {\bibfnamefont {J.~A.}\ \bibnamefont {Jaramillo-Villegas}}, \bibinfo {author} {\bibfnamefont {D.~E.}\ \bibnamefont {Leaird}},\ and\ \bibinfo {author} {\bibfnamefont {A.~M.}\ \bibnamefont {Weiner}},\ }\bibfield  {title} {\bibinfo {title} {Characterization of coherent quantum frequency combs using electro-optic phase modulation},\ }\href {https://doi.org/10.1103/PhysRevA.97.013813} {\bibfield  {journal} {\bibinfo  {journal} {Phys. Rev. A}\ }\textbf {\bibinfo {volume} {97}},\ \bibinfo {pages} {013813} (\bibinfo {year} {2018})}\BibitemShut {NoStop}%
\bibitem [{\citenamefont {Thew}\ \emph {et~al.}(2002)\citenamefont {Thew}, \citenamefont {Nemoto}, \citenamefont {White},\ and\ \citenamefont {Munro}}]{thew}%
  \BibitemOpen
  \bibfield  {author} {\bibinfo {author} {\bibfnamefont {R.~T.}\ \bibnamefont {Thew}}, \bibinfo {author} {\bibfnamefont {K.}~\bibnamefont {Nemoto}}, \bibinfo {author} {\bibfnamefont {A.~G.}\ \bibnamefont {White}},\ and\ \bibinfo {author} {\bibfnamefont {W.~J.}\ \bibnamefont {Munro}},\ }\bibfield  {title} {\bibinfo {title} {Qudit quantum-state tomography},\ }\href {https://doi.org/10.1103/PhysRevA.66.012303} {\bibfield  {journal} {\bibinfo  {journal} {Phys. Rev. A}\ }\textbf {\bibinfo {volume} {66}},\ \bibinfo {pages} {012303} (\bibinfo {year} {2002})}\BibitemShut {NoStop}%
\bibitem [{\citenamefont {James}\ \emph {et~al.}(2001)\citenamefont {James}, \citenamefont {Kwiat}, \citenamefont {Munro},\ and\ \citenamefont {White}}]{dfvj}%
  \BibitemOpen
  \bibfield  {author} {\bibinfo {author} {\bibfnamefont {D.~F.~V.}\ \bibnamefont {James}}, \bibinfo {author} {\bibfnamefont {P.~G.}\ \bibnamefont {Kwiat}}, \bibinfo {author} {\bibfnamefont {W.~J.}\ \bibnamefont {Munro}},\ and\ \bibinfo {author} {\bibfnamefont {A.~G.}\ \bibnamefont {White}},\ }\bibfield  {title} {\bibinfo {title} {Measurement of qubits},\ }\href {https://doi.org/10.1103/PhysRevA.64.052312} {\bibfield  {journal} {\bibinfo  {journal} {Phys. Rev. A}\ }\textbf {\bibinfo {volume} {64}},\ \bibinfo {pages} {052312} (\bibinfo {year} {2001})}\BibitemShut {NoStop}%
\bibitem [{\citenamefont {Lukens}\ \emph {et~al.}(2020{\natexlab{b}})\citenamefont {Lukens}, \citenamefont {Law}, \citenamefont {Jasra},\ and\ \citenamefont {Lougovski}}]{lukens2}%
  \BibitemOpen
  \bibfield  {author} {\bibinfo {author} {\bibfnamefont {J.~M.}\ \bibnamefont {Lukens}}, \bibinfo {author} {\bibfnamefont {K.~J.}\ \bibnamefont {Law}}, \bibinfo {author} {\bibfnamefont {A.}~\bibnamefont {Jasra}},\ and\ \bibinfo {author} {\bibfnamefont {P.}~\bibnamefont {Lougovski}},\ }\bibfield  {title} {\bibinfo {title} {A practical and efficient approach for bayesian quantum state estimation},\ }\href {https://doi.org/10.1088/1367-2630/ab8efa} {\bibfield  {journal} {\bibinfo  {journal} {New Journal of Physics}\ }\textbf {\bibinfo {volume} {22}},\ \bibinfo {pages} {063038} (\bibinfo {year} {2020}{\natexlab{b}})}\BibitemShut {NoStop}%
\bibitem [{\citenamefont {Lohani}\ \emph {et~al.}(2020)\citenamefont {Lohani}, \citenamefont {Kirby}, \citenamefont {Brodsky}, \citenamefont {Danaci},\ and\ \citenamefont {Glasser}}]{lohani}%
  \BibitemOpen
  \bibfield  {author} {\bibinfo {author} {\bibfnamefont {S.}~\bibnamefont {Lohani}}, \bibinfo {author} {\bibfnamefont {B.~T.}\ \bibnamefont {Kirby}}, \bibinfo {author} {\bibfnamefont {M.}~\bibnamefont {Brodsky}}, \bibinfo {author} {\bibfnamefont {O.}~\bibnamefont {Danaci}},\ and\ \bibinfo {author} {\bibfnamefont {R.~T.}\ \bibnamefont {Glasser}},\ }\bibfield  {title} {\bibinfo {title} {Machine learning assisted quantum state estimation},\ }\href {https://doi.org/10.1088/2632-2153/ab9a21} {\bibfield  {journal} {\bibinfo  {journal} {Machine Learning: Science and Technology}\ }\textbf {\bibinfo {volume} {1}},\ \bibinfo {pages} {035007} (\bibinfo {year} {2020})}\BibitemShut {NoStop}%
\bibitem [{\citenamefont {Danaci}\ \emph {et~al.}(2021)\citenamefont {Danaci}, \citenamefont {Lohani}, \citenamefont {Kirby},\ and\ \citenamefont {Glasser}}]{danaci}%
  \BibitemOpen
  \bibfield  {author} {\bibinfo {author} {\bibfnamefont {O.}~\bibnamefont {Danaci}}, \bibinfo {author} {\bibfnamefont {S.}~\bibnamefont {Lohani}}, \bibinfo {author} {\bibfnamefont {B.~T.}\ \bibnamefont {Kirby}},\ and\ \bibinfo {author} {\bibfnamefont {R.~T.}\ \bibnamefont {Glasser}},\ }\bibfield  {title} {\bibinfo {title} {Machine learning pipeline for quantum state estimation with incomplete measurements},\ }\href {https://doi.org/10.1088/2632-2153/abe5f5} {\bibfield  {journal} {\bibinfo  {journal} {Machine Learning: Science and Technology}\ }\textbf {\bibinfo {volume} {2}},\ \bibinfo {pages} {035014} (\bibinfo {year} {2021})}\BibitemShut {NoStop}%
\bibitem [{\citenamefont {Lohani}\ \emph {et~al.}(2023)\citenamefont {Lohani}, \citenamefont {Lukens}, \citenamefont {Davis}, \citenamefont {Khannejad}, \citenamefont {Regmi}, \citenamefont {Jones}, \citenamefont {Glasser}, \citenamefont {Searles},\ and\ \citenamefont {Kirby}}]{lohani_2023}%
  \BibitemOpen
  \bibfield  {author} {\bibinfo {author} {\bibfnamefont {S.}~\bibnamefont {Lohani}}, \bibinfo {author} {\bibfnamefont {J.~M.}\ \bibnamefont {Lukens}}, \bibinfo {author} {\bibfnamefont {A.~A.}\ \bibnamefont {Davis}}, \bibinfo {author} {\bibfnamefont {A.}~\bibnamefont {Khannejad}}, \bibinfo {author} {\bibfnamefont {S.}~\bibnamefont {Regmi}}, \bibinfo {author} {\bibfnamefont {D.~E.}\ \bibnamefont {Jones}}, \bibinfo {author} {\bibfnamefont {R.~T.}\ \bibnamefont {Glasser}}, \bibinfo {author} {\bibfnamefont {T.~A.}\ \bibnamefont {Searles}},\ and\ \bibinfo {author} {\bibfnamefont {B.~T.}\ \bibnamefont {Kirby}},\ }\bibfield  {title} {\bibinfo {title} {Demonstration of machine-learning-enhanced bayesian quantum state estimation},\ }\href {https://doi.org/10.1088/1367-2630/ace6c8} {\bibfield  {journal} {\bibinfo  {journal} {IOP Publishing}\ }\textbf {\bibinfo {volume} {25}},\ \bibinfo {pages} {083009} (\bibinfo {year} {2023})}\BibitemShut {NoStop}%
\bibitem [{\citenamefont {Cha}\ \emph {et~al.}(2021)\citenamefont {Cha}, \citenamefont {Ginsparg}, \citenamefont {Wu}, \citenamefont {Carrasquilla}, \citenamefont {McMahon},\ and\ \citenamefont {Kim}}]{transformers}%
  \BibitemOpen
  \bibfield  {author} {\bibinfo {author} {\bibfnamefont {P.}~\bibnamefont {Cha}}, \bibinfo {author} {\bibfnamefont {P.}~\bibnamefont {Ginsparg}}, \bibinfo {author} {\bibfnamefont {F.}~\bibnamefont {Wu}}, \bibinfo {author} {\bibfnamefont {J.}~\bibnamefont {Carrasquilla}}, \bibinfo {author} {\bibfnamefont {P.~L.}\ \bibnamefont {McMahon}},\ and\ \bibinfo {author} {\bibfnamefont {E.-A.}\ \bibnamefont {Kim}},\ }\bibfield  {title} {\bibinfo {title} {Attention-based quantum tomography},\ }\bibfield  {journal} {\bibinfo  {journal} {Machine Learning: Science and Technology}\ }\textbf {\bibinfo {volume} {3}},\ \href {https://doi.org/10.1088/2632-2153/ac362b} {10.1088/2632-2153/ac362b} (\bibinfo {year} {2021})\BibitemShut {NoStop}%
\bibitem [{\citenamefont {Ma}\ and\ \citenamefont {Yung}(2018)}]{ma}%
  \BibitemOpen
  \bibfield  {author} {\bibinfo {author} {\bibfnamefont {Y.-C.}\ \bibnamefont {Ma}}\ and\ \bibinfo {author} {\bibfnamefont {M.-H.}\ \bibnamefont {Yung}},\ }\bibfield  {title} {\bibinfo {title} {Transforming bell’s inequalities into state classifiers with machine learning},\ }\bibfield  {journal} {\bibinfo  {journal} {npj Quantum Information}\ }\textbf {\bibinfo {volume} {4}},\ \href {https://doi.org/10.1038/s41534-018-0081-3} {10.1038/s41534-018-0081-3} (\bibinfo {year} {2018})\BibitemShut {NoStop}%
\bibitem [{\citenamefont {Lu}\ \emph {et~al.}(2018)\citenamefont {Lu}, \citenamefont {Huang}, \citenamefont {Li}, \citenamefont {Li}, \citenamefont {Chen}, \citenamefont {Lu}, \citenamefont {Ji}, \citenamefont {Shen}, \citenamefont {Zhou},\ and\ \citenamefont {Zeng}}]{lu}%
  \BibitemOpen
  \bibfield  {author} {\bibinfo {author} {\bibfnamefont {S.}~\bibnamefont {Lu}}, \bibinfo {author} {\bibfnamefont {S.}~\bibnamefont {Huang}}, \bibinfo {author} {\bibfnamefont {K.}~\bibnamefont {Li}}, \bibinfo {author} {\bibfnamefont {J.}~\bibnamefont {Li}}, \bibinfo {author} {\bibfnamefont {J.}~\bibnamefont {Chen}}, \bibinfo {author} {\bibfnamefont {D.}~\bibnamefont {Lu}}, \bibinfo {author} {\bibfnamefont {Z.}~\bibnamefont {Ji}}, \bibinfo {author} {\bibfnamefont {Y.}~\bibnamefont {Shen}}, \bibinfo {author} {\bibfnamefont {D.}~\bibnamefont {Zhou}},\ and\ \bibinfo {author} {\bibfnamefont {B.}~\bibnamefont {Zeng}},\ }\bibfield  {title} {\bibinfo {title} {Separability-entanglement classifier via machine learning},\ }\href {https://doi.org/10.1103/PhysRevA.98.012315} {\bibfield  {journal} {\bibinfo  {journal} {Phys. Rev. A}\ }\textbf {\bibinfo {volume} {98}},\ \bibinfo {pages} {012315} (\bibinfo {year} {2018})}\BibitemShut {NoStop}%
\bibitem [{\citenamefont {Greenwood}\ \emph {et~al.}(2023)\citenamefont {Greenwood}, \citenamefont {Wu}, \citenamefont {Zhu}, \citenamefont {Kirby},\ and\ \citenamefont {Qian}}]{PhysRevApplied.19.034058}%
  \BibitemOpen
  \bibfield  {author} {\bibinfo {author} {\bibfnamefont {A.~C.}\ \bibnamefont {Greenwood}}, \bibinfo {author} {\bibfnamefont {L.~T.}\ \bibnamefont {Wu}}, \bibinfo {author} {\bibfnamefont {E.~Y.}\ \bibnamefont {Zhu}}, \bibinfo {author} {\bibfnamefont {B.~T.}\ \bibnamefont {Kirby}},\ and\ \bibinfo {author} {\bibfnamefont {L.}~\bibnamefont {Qian}},\ }\bibfield  {title} {\bibinfo {title} {Machine-learning-derived entanglement witnesses},\ }\href {https://doi.org/10.1103/PhysRevApplied.19.034058} {\bibfield  {journal} {\bibinfo  {journal} {Phys. Rev. Appl.}\ }\textbf {\bibinfo {volume} {19}},\ \bibinfo {pages} {034058} (\bibinfo {year} {2023})}\BibitemShut {NoStop}%
\bibitem [{\citenamefont {T\'oth}\ and\ \citenamefont {G\"uhne}(2005)}]{toth}%
  \BibitemOpen
  \bibfield  {author} {\bibinfo {author} {\bibfnamefont {G.}~\bibnamefont {T\'oth}}\ and\ \bibinfo {author} {\bibfnamefont {O.}~\bibnamefont {G\"uhne}},\ }\bibfield  {title} {\bibinfo {title} {Entanglement detection in the stabilizer formalism},\ }\href {https://doi.org/10.1103/PhysRevA.72.022340} {\bibfield  {journal} {\bibinfo  {journal} {Phys. Rev. A}\ }\textbf {\bibinfo {volume} {72}},\ \bibinfo {pages} {022340} (\bibinfo {year} {2005})}\BibitemShut {NoStop}%
\bibitem [{\citenamefont {Gühne}\ and\ \citenamefont {Tóth}(2009)}]{guhne}%
  \BibitemOpen
  \bibfield  {author} {\bibinfo {author} {\bibfnamefont {O.}~\bibnamefont {Gühne}}\ and\ \bibinfo {author} {\bibfnamefont {G.}~\bibnamefont {Tóth}},\ }\bibfield  {title} {\bibinfo {title} {Entanglement detection},\ }\href {https://doi.org/10.1016/j.physrep.2009.02.004} {\bibfield  {journal} {\bibinfo  {journal} {Physics Reports}\ }\textbf {\bibinfo {volume} {474}},\ \bibinfo {pages} {1–75} (\bibinfo {year} {2009})}\BibitemShut {NoStop}%
\bibitem [{\citenamefont {Campbell}\ and\ \citenamefont {Ying}(2011)}]{svm_textbook}%
  \BibitemOpen
  \bibfield  {author} {\bibinfo {author} {\bibfnamefont {C.}~\bibnamefont {Campbell}}\ and\ \bibinfo {author} {\bibfnamefont {Y.}~\bibnamefont {Ying}},\ }\bibinfo {title} {Support vector machines for classification},\ in\ \href {https://doi.org/10.1007/978-3-031-01552-6_1} {\emph {\bibinfo {booktitle} {Learning with Support Vector Machines}}}\ (\bibinfo  {publisher} {Springer International Publishing},\ \bibinfo {address} {Cham},\ \bibinfo {year} {2011})\ pp.\ \bibinfo {pages} {1--25}\BibitemShut {NoStop}%
\bibitem [{\citenamefont {Boyd}\ and\ \citenamefont {Vandenberghe}(2023)}]{Boyd}%
  \BibitemOpen
  \bibfield  {author} {\bibinfo {author} {\bibfnamefont {S.~P.}\ \bibnamefont {Boyd}}\ and\ \bibinfo {author} {\bibfnamefont {L.}~\bibnamefont {Vandenberghe}},\ }\href@noop {} {\emph {\bibinfo {title} {Convex optimization}}}\ (\bibinfo  {publisher} {Cambridge University Press},\ \bibinfo {year} {2023})\BibitemShut {NoStop}%
\bibitem [{\citenamefont {Iranmehr}\ \emph {et~al.}(2019)\citenamefont {Iranmehr}, \citenamefont {Masnadi-Shirazi},\ and\ \citenamefont {Vasconcelos}}]{cost_SVM}%
  \BibitemOpen
  \bibfield  {author} {\bibinfo {author} {\bibfnamefont {A.}~\bibnamefont {Iranmehr}}, \bibinfo {author} {\bibfnamefont {H.}~\bibnamefont {Masnadi-Shirazi}},\ and\ \bibinfo {author} {\bibfnamefont {N.}~\bibnamefont {Vasconcelos}},\ }\bibfield  {title} {\bibinfo {title} {Cost-sensitive support vector machines},\ }\href {https://doi.org/https://doi.org/10.1016/j.neucom.2018.11.099} {\bibfield  {journal} {\bibinfo  {journal} {Neurocomputing}\ }\textbf {\bibinfo {volume} {343}},\ \bibinfo {pages} {50} (\bibinfo {year} {2019})},\ \bibinfo {note} {learning in the Presence of Class Imbalance and Concept Drift}\BibitemShut {NoStop}%
\bibitem [{\citenamefont {Yang}\ \emph {et~al.}(2005)\citenamefont {Yang}, \citenamefont {Song},\ and\ \citenamefont {Cao}}]{weight_SVM}%
  \BibitemOpen
  \bibfield  {author} {\bibinfo {author} {\bibfnamefont {X.}~\bibnamefont {Yang}}, \bibinfo {author} {\bibfnamefont {Q.}~\bibnamefont {Song}},\ and\ \bibinfo {author} {\bibfnamefont {A.}~\bibnamefont {Cao}},\ }\bibfield  {title} {\bibinfo {title} {Weighted support vector machine for data classification},\ }in\ \href {https://doi.org/10.1109/IJCNN.2005.1555965} {\emph {\bibinfo {booktitle} {Proceedings. 2005 IEEE International Joint Conference on Neural Networks, 2005.}}},\ Vol.~\bibinfo {volume} {2}\ (\bibinfo {year} {2005})\ pp.\ \bibinfo {pages} {859--864 vol. 2}\BibitemShut {NoStop}%
\bibitem [{\citenamefont {Jungnitsch}\ \emph {et~al.}(2011)\citenamefont {Jungnitsch}, \citenamefont {Moroder},\ and\ \citenamefont {G\"uhne}}]{taming}%
  \BibitemOpen
  \bibfield  {author} {\bibinfo {author} {\bibfnamefont {B.}~\bibnamefont {Jungnitsch}}, \bibinfo {author} {\bibfnamefont {T.}~\bibnamefont {Moroder}},\ and\ \bibinfo {author} {\bibfnamefont {O.}~\bibnamefont {G\"uhne}},\ }\bibfield  {title} {\bibinfo {title} {Taming multiparticle entanglement},\ }\href {https://doi.org/10.1103/PhysRevLett.106.190502} {\bibfield  {journal} {\bibinfo  {journal} {Phys. Rev. Lett.}\ }\textbf {\bibinfo {volume} {106}},\ \bibinfo {pages} {190502} (\bibinfo {year} {2011})}\BibitemShut {NoStop}%
\bibitem [{\citenamefont {Lohani}\ \emph {et~al.}(2022)\citenamefont {Lohani}, \citenamefont {Lukens}, \citenamefont {Glasser}, \citenamefont {Searles},\ and\ \citenamefont {Kirby}}]{lohani2022data}%
  \BibitemOpen
  \bibfield  {author} {\bibinfo {author} {\bibfnamefont {S.}~\bibnamefont {Lohani}}, \bibinfo {author} {\bibfnamefont {J.~M.}\ \bibnamefont {Lukens}}, \bibinfo {author} {\bibfnamefont {R.~T.}\ \bibnamefont {Glasser}}, \bibinfo {author} {\bibfnamefont {T.~A.}\ \bibnamefont {Searles}},\ and\ \bibinfo {author} {\bibfnamefont {B.~T.}\ \bibnamefont {Kirby}},\ }\bibfield  {title} {\bibinfo {title} {Data-centric machine learning in quantum information science},\ }\href@noop {} {\bibfield  {journal} {\bibinfo  {journal} {Machine Learning: Science and Technology}\ }\textbf {\bibinfo {volume} {3}},\ \bibinfo {pages} {04LT01} (\bibinfo {year} {2022})}\BibitemShut {NoStop}%
\bibitem [{\citenamefont {Wang}\ \emph {et~al.}(2014)\citenamefont {Wang}, \citenamefont {Li}, \citenamefont {Liu}, \citenamefont {Zhang},\ and\ \citenamefont {Zhang}}]{Wang}%
  \BibitemOpen
  \bibfield  {author} {\bibinfo {author} {\bibfnamefont {S.}~\bibnamefont {Wang}}, \bibinfo {author} {\bibfnamefont {Z.}~\bibnamefont {Li}}, \bibinfo {author} {\bibfnamefont {C.}~\bibnamefont {Liu}}, \bibinfo {author} {\bibfnamefont {X.}~\bibnamefont {Zhang}},\ and\ \bibinfo {author} {\bibfnamefont {H.}~\bibnamefont {Zhang}},\ }\bibfield  {title} {\bibinfo {title} {Training data reduction to speed up svm training},\ }\href@noop {} {\bibfield  {journal} {\bibinfo  {journal} {Applied Intelligence}\ }\textbf {\bibinfo {volume} {41}},\ \bibinfo {pages} {405} (\bibinfo {year} {2014})}\BibitemShut {NoStop}%
\bibitem [{\citenamefont {Nan}\ \emph {et~al.}(2017)\citenamefont {Nan}, \citenamefont {Sun}, \citenamefont {Chen}, \citenamefont {Lin},\ and\ \citenamefont {Toh}}]{Nan}%
  \BibitemOpen
  \bibfield  {author} {\bibinfo {author} {\bibfnamefont {S.}~\bibnamefont {Nan}}, \bibinfo {author} {\bibfnamefont {L.}~\bibnamefont {Sun}}, \bibinfo {author} {\bibfnamefont {B.}~\bibnamefont {Chen}}, \bibinfo {author} {\bibfnamefont {Z.}~\bibnamefont {Lin}},\ and\ \bibinfo {author} {\bibfnamefont {K.-A.}\ \bibnamefont {Toh}},\ }\bibfield  {title} {\bibinfo {title} {Density-dependent quantized least squares support vector machine for large data sets},\ }\href {https://doi.org/10.1109/TNNLS.2015.2504382} {\bibfield  {journal} {\bibinfo  {journal} {IEEE Transactions on Neural Networks and Learning Systems}\ }\textbf {\bibinfo {volume} {28}},\ \bibinfo {pages} {94} (\bibinfo {year} {2017})}\BibitemShut {NoStop}%
\bibitem [{Ten()}]{TensorFlow}%
  \BibitemOpen
  \href {https://www.tensorflow.org/api_docs/python/tf/keras/optimizers/Adam} {}\BibitemShut {NoStop}%
\bibitem [{\citenamefont {Zyczkowski}\ and\ \citenamefont {Kus}(1994)}]{haar}%
  \BibitemOpen
  \bibfield  {author} {\bibinfo {author} {\bibfnamefont {K.}~\bibnamefont {Zyczkowski}}\ and\ \bibinfo {author} {\bibfnamefont {M.}~\bibnamefont {Kus}},\ }\bibfield  {title} {\bibinfo {title} {Random unitary matrices},\ }\href {https://doi.org/10.1088/0305-4470/27/12/028} {\bibfield  {journal} {\bibinfo  {journal} {Journal of Physics A: Mathematical and General}\ }\textbf {\bibinfo {volume} {27}},\ \bibinfo {pages} {4235} (\bibinfo {year} {1994})}\BibitemShut {NoStop}%
\bibitem [{\citenamefont {Vintskevich}\ \emph {et~al.}(2023)\citenamefont {Vintskevich}, \citenamefont {Bao}, \citenamefont {Nomerotski}, \citenamefont {Stankus},\ and\ \citenamefont {Grigoriev}}]{new}%
  \BibitemOpen
  \bibfield  {author} {\bibinfo {author} {\bibfnamefont {S.~V.}\ \bibnamefont {Vintskevich}}, \bibinfo {author} {\bibfnamefont {N.}~\bibnamefont {Bao}}, \bibinfo {author} {\bibfnamefont {A.}~\bibnamefont {Nomerotski}}, \bibinfo {author} {\bibfnamefont {P.}~\bibnamefont {Stankus}},\ and\ \bibinfo {author} {\bibfnamefont {D.~A.}\ \bibnamefont {Grigoriev}},\ }\bibfield  {title} {\bibinfo {title} {Classification of four-qubit entangled states via machine learning},\ }\href {https://doi.org/10.1103/PhysRevA.107.032421} {\bibfield  {journal} {\bibinfo  {journal} {Phys. Rev. A}\ }\textbf {\bibinfo {volume} {107}},\ \bibinfo {pages} {032421} (\bibinfo {year} {2023})}\BibitemShut {NoStop}%
\bibitem [{\citenamefont {Verstraete}\ \emph {et~al.}(2002)\citenamefont {Verstraete}, \citenamefont {Dehaene}, \citenamefont {De~Moor},\ and\ \citenamefont {Verschelde}}]{new_4qubit_ref}%
  \BibitemOpen
  \bibfield  {author} {\bibinfo {author} {\bibfnamefont {F.}~\bibnamefont {Verstraete}}, \bibinfo {author} {\bibfnamefont {J.}~\bibnamefont {Dehaene}}, \bibinfo {author} {\bibfnamefont {B.}~\bibnamefont {De~Moor}},\ and\ \bibinfo {author} {\bibfnamefont {H.}~\bibnamefont {Verschelde}},\ }\bibfield  {title} {\bibinfo {title} {Four qubits can be entangled in nine different ways},\ }\href {https://doi.org/10.1103/PhysRevA.65.052112} {\bibfield  {journal} {\bibinfo  {journal} {Phys. Rev. A}\ }\textbf {\bibinfo {volume} {65}},\ \bibinfo {pages} {052112} (\bibinfo {year} {2002})}\BibitemShut {NoStop}%
\bibitem [{\citenamefont {Baydin}\ \emph {et~al.}(2018)\citenamefont {Baydin}, \citenamefont {Pearlmutter}, \citenamefont {Radul},\ and\ \citenamefont {Siskind}}]{diff}%
  \BibitemOpen
  \bibfield  {author} {\bibinfo {author} {\bibfnamefont {A.~G.}\ \bibnamefont {Baydin}}, \bibinfo {author} {\bibfnamefont {B.~A.}\ \bibnamefont {Pearlmutter}}, \bibinfo {author} {\bibfnamefont {A.~A.}\ \bibnamefont {Radul}},\ and\ \bibinfo {author} {\bibfnamefont {J.~M.}\ \bibnamefont {Siskind}},\ }\bibfield  {title} {\bibinfo {title} {Automatic differentiation in machine learning: a survey},\ }\href {http://jmlr.org/papers/v18/17-468.html} {\bibfield  {journal} {\bibinfo  {journal} {Journal of Machine Learning Research}\ }\textbf {\bibinfo {volume} {18}},\ \bibinfo {pages} {1} (\bibinfo {year} {2018})}\BibitemShut {NoStop}%
\bibitem [{\citenamefont {Andrews}(1984)}]{partitions}%
  \BibitemOpen
  \bibfield  {author} {\bibinfo {author} {\bibfnamefont {G.~E.}\ \bibnamefont {Andrews}},\ }\bibinfo {title} {The hardy–ramanujan–rademacher expansion of p(n)},\ in\ \href {https://doi.org/10.1017/CBO9780511608650.008} {\emph {\bibinfo {booktitle} {The Theory of Partitions}}},\ \bibinfo {series and number} {Encyclopedia of Mathematics and its Applications}\ (\bibinfo  {publisher} {Cambridge University Press},\ \bibinfo {year} {1984})\ p.\ \bibinfo {pages} {68–87}\BibitemShut {NoStop}%
\bibitem [{\citenamefont {Bourennane}\ \emph {et~al.}(2004)\citenamefont {Bourennane}, \citenamefont {Eibl}, \citenamefont {Kurtsiefer}, \citenamefont {Gaertner}, \citenamefont {Weinfurter}, \citenamefont {G\"uhne}, \citenamefont {Hyllus}, \citenamefont {Bru\ss{}}, \citenamefont {Lewenstein},\ and\ \citenamefont {Sanpera}}]{fidelity}%
  \BibitemOpen
  \bibfield  {author} {\bibinfo {author} {\bibfnamefont {M.}~\bibnamefont {Bourennane}}, \bibinfo {author} {\bibfnamefont {M.}~\bibnamefont {Eibl}}, \bibinfo {author} {\bibfnamefont {C.}~\bibnamefont {Kurtsiefer}}, \bibinfo {author} {\bibfnamefont {S.}~\bibnamefont {Gaertner}}, \bibinfo {author} {\bibfnamefont {H.}~\bibnamefont {Weinfurter}}, \bibinfo {author} {\bibfnamefont {O.}~\bibnamefont {G\"uhne}}, \bibinfo {author} {\bibfnamefont {P.}~\bibnamefont {Hyllus}}, \bibinfo {author} {\bibfnamefont {D.}~\bibnamefont {Bru\ss{}}}, \bibinfo {author} {\bibfnamefont {M.}~\bibnamefont {Lewenstein}},\ and\ \bibinfo {author} {\bibfnamefont {A.}~\bibnamefont {Sanpera}},\ }\bibfield  {title} {\bibinfo {title} {Experimental detection of multipartite entanglement using witness operators},\ }\href {https://doi.org/10.1103/PhysRevLett.92.087902} {\bibfield  {journal} {\bibinfo  {journal} {Phys. Rev. Lett.}\ }\textbf {\bibinfo {volume} {92}},\ \bibinfo {pages} {087902} (\bibinfo {year} {2004})}\BibitemShut {NoStop}%
\bibitem [{\citenamefont {Kingma}\ and\ \citenamefont {Ba}(2014)}]{adam}%
  \BibitemOpen
  \bibfield  {author} {\bibinfo {author} {\bibfnamefont {D.~P.}\ \bibnamefont {Kingma}}\ and\ \bibinfo {author} {\bibfnamefont {J.}~\bibnamefont {Ba}},\ }\bibfield  {title} {\bibinfo {title} {Adam: A method for stochastic optimization},\ }\href {https://api.semanticscholar.org/CorpusID:6628106} {\bibfield  {journal} {\bibinfo  {journal} {CoRR}\ }\textbf {\bibinfo {volume} {abs/1412.6980}} (\bibinfo {year} {2014})}\BibitemShut {NoStop}%
\end{thebibliography}%

\appendix
\section{\label{sec:gradient_descent} Gradient Descent}

The method of gradient descent is a means of numerically minimizing a loss function over some number of discrete iterations. In each
iteration, the gradient of the loss function is calculated across all
free variables, which describes how each variable should be changed to
minimize the loss function the most for that step.

This method is most simply implemented for unconstrained optimization
problems, where each variable can be freely incremented according to the
gradient of the loss function. As we discuss below it is useful to
re-parameterize a constrained optimization problem into one which is
unconstrained, as then powerful optimizers used for machine learning
which implement this method can be taken advantage of.

We used the popular ADAM optimizer \cite{adam} for the purposes of this project. It can efficiently solve problems
with many parameters which will become more relevant if this method is extended beyond the results here. As we detail in my Methods and Results section, we
used gradient descent and the measures of similarity between quantum
states to find support vectors for training entanglement witnesses.

\section{\label{sec:mso_parameterization} Differential Mixed State Optimization Parameterization Scheme}
K-separable pure states are defined as a product state where an \(N\) qubit state has \(K\) groups of entangled qubits, with \(K \in [1,N]\). K-separable pure states then have at least one and at most N qubits which are not entangled with the others. In the case of 3 qubits, a biseparable pure state across qubits 1, 2, and 3 can be written as \(|\Psi_{1|23} \rangle = |\phi_1\rangle \otimes |\phi_{23}\rangle \) or \(|\Psi_{2|13} \rangle = |\phi_2\rangle \otimes |\phi_{13}\rangle \) or \(|\Psi_{3|12} \rangle = |\phi_{12}\rangle \otimes |\phi_3\rangle \) where \(|\phi_{AB}\rangle\) refers to a bipartite state across qubits A and B which may be entangled \cite{taming}. Similar reasoning applies for systems of more qubits, where k-separable states must be able to be written in terms of at least one tensor product and necessary swapping operators.

Now a k-separable mixed state of
\(N\) qubits
can be represented \cite{taming} as
\begin{equation} \label{eq: mixed} \rho = \sum_{i}^A p_i |\psi_i\rangle \langle \psi_i|\end{equation}
where each \(|\psi_i\rangle \)
corresponds to a k-separable pure state with \( p_i \in [0,1] \)
and \(\sum p_i = 1\). \(A\) is chosen such that we can guarantee that both pure states and the maximally mixed state can be represented across all possible combinations of subspaces. This reasoning and our method of choosing \(A\) is defined later in this section.

Each pure state can be further partitioned into
\(n\)
arbitrary pure states of \(\nu_i\)
qubits each where 
\begin{equation} \label{eq: num_partitions} \sum_i^n \nu_i = N.\end{equation}

Finally, an arbitrary pure state of \( \nu\)
qubits can be represented as
\begin{equation} \label{eq: pure_state_parameterization}  |\phi \rangle = 
\left[ \begin{matrix}
k_0
\\k_1e^{i\theta_1}
\\k_2e^{i\theta_2}
\\\vdots
\\k_{(2^{\nu})-1}e^{i\theta_{(2^{\nu})-1}}
\end{matrix} \right]\end{equation}

Here, the magnitudes \(k_i \in [0,1]\) and \(\sum_i k_i ^2 = 1\), while each
\(\theta_i \) can effectively vary freely for the purposes of optimization.

Now, Tensorflow is a powerful tool for \textit{unconstrained}
optimization, which means all parameters passed to the optimizer must be
allowed to vary freely. To accommodate the constraints on \(k_i \) and \(p_i\) it is necessary to create new constrained parameters by manipulating free parameters.

Suppose
\(\{x_0, x_1,\dots x_n\}\)
are freely varying parameters initialized not all at zero. Then we set
\begin{equation} \label{eq: constrain_k}k_i = \frac{x^2_i}{\sqrt{\sum_{j = 0}^{n} x^4_j}}\end{equation} 
to force the constraints on k to be fulfilled. Similarly for some additional set of free parameters \(\{y_0, y_1,\dots y_n\}\)
choosing each
\(p_i\)
as
\begin{equation} \label{eq: constrain_p} p_i = \frac{y_i^2}{\sum_j y_j^2}\end{equation}

ensures the values of 
\(p\) are positive and sum to 1.

Note that we considered partitioning each pure state first into groups of at least two qubits, with one additional separable single qubit subsystem as needed to reach a state of 4 or 5 qubits in total. As we fully parameterize the space of larger subsystems, with certain choices of parameters, those subsystems could represent states which are themselves separable, meaning we don't have to consider partitions explicitly involving more than one single qubit state. We are forced to parameterize at least one subsystem of at least one qubit separately to guarantee the final overall state is always separable.

To completely span the space of mixed k-separable states, each possible
permutation of pure k-separable states must be represented in the set
\(\{|\phi_j\rangle\}\).
We found that the optimizer makes correct adjustments to bias terms
in GHZ witnesses when each permutation of each pure state is represented
by some number of independent copies of the state, according to a rule
specified as follows: If

\begin{equation} \label{eq: pure_factoring} |\psi_i\rangle = \bigotimes _j|\phi _j\rangle \end{equation} 

and each \( |\phi_j \rangle \) is an arbitrary pure state of
\(\nu_j\) qubits, then we fully parameterize \(\alpha\)
independent pure states representing the same partitioning and permutations of subspaces as \(|\phi_j\rangle\), where

\begin{equation} \label{eq: alpha} \alpha = \max_j 2^{v_j} \end{equation} 

For example, if we were to consider a 3 qubit pure state with subsystem 1 of qubits 1 and 2, which are possibly entangled, and subsystem 2 of qubit 3 which is always separable. We would have \(v_1 = 2\) for the subsystem of entangled qubits 1 and 2 and \(v_2 = 1\) for the remaining separable qubit. We have \(\alpha = \max_j {2^{v_j}} = \max{2^1, 2^2} = 2^2 = 4\), so we would represent \(2^2 = 4\) unique states, which can vary independently, with this arrangement of qubits.

Since each group of qubits exists in different subspaces, a convex combination of density operators can be thought of as creating mixed states in each subspace independently. To reach the maximally mixed state in each subspace, we need to sum at least as many density operators as there are basis states in that subspace. By combining as many freely varying pure states as there are basis states in a subspace, we supposed that any mixed state could be reached in the convex sum. Then for a system comprised of several subsystems, we create as many independently varying pure states in the whole system as there are basis states in the largest subsystem, which ensures the largest subsystem meets this requirement and the smaller subsystems have additional redundancy. 

With the parameterization procedure described above, we can guarantee that we can produce, in any subsystem, completely pure states (with only one pure state contributing to the sum) and maximally mixed states (with every pure state contributing to the sum). Then, by introducing redundancy in subsystem parameterization, we can also produce states with varying levels of purity for the whole system.  As shown in the case of GHZ witness we produced, our optimization using the above-mentioned parameterization seems adequate for the purpose of adjusting the bias of the witness. Therefore, we expect the optimization to work well for other witnesses. Additionally, since we exhaustively consider all possible groupings of entangled qubits (through partitions and permutations), we believe we can represent any mixed state represented by Equation \ref{eq: mixed}.

To find all permutations for any number of qubits, we considered how
many ways the set of qubits can be arranged into groups of at least two.
For three qubits, there are only three permutations; where qubits 1 and 2 are
allowed to be entangled, qubits 2 and 3 are allowed to be entangled, and
qubits 1 and 3 are allowed to be entangled. Denote these permutations as
\(3|12\), \(1|23\), and \(2|13\), respectively, where numbers not separated by the | symbol label entangled qubits.

To reach different permutations, a swap operator (\(S\)) can be applied. Also let \(|\phi_1 \rangle\) be a single qubit pure state and

A \(1|23\)
state can then be parameterized as

\begin{equation} \label{eq: 1|23} |\psi_{1|23}\rangle =|\phi_1\rangle \otimes |\phi_{23}\rangle\end{equation}

while a \(2|13\)
state could be parameterized as
\begin{equation} \label{eq: 2|13} |\psi_{2|13}\rangle =(S\otimes I)|\psi _{1|23}\rangle\end{equation}
and a \(3|12\) state could be parameterized as

\begin{equation} \label{eq: 3|12} |\psi_{3|12}\rangle =|\phi_2\rangle \otimes |\phi_1\rangle \end{equation}.

For an
\(N\)
qubit pure state, to swap qubits
\(\nu \) and \(\nu + 1\),
a swap operator can be constructed as

\begin{equation} \label{eq: swap_adjacent_qubits} S_\nu^N = I^{\otimes (\nu -1)}\otimes S\otimes I^{\otimes (N - \nu - 1)}\end{equation} .

An operator switching the states of arbitrary qubits
\(a\) and \(b\) in an \(N\)
qubit pure state can then be determined as 
\begin{equation} \label{eq: swap_any_qubits} S_{ab}^N= \prod_{\nu=a}^{b-1}S_{\nu}^N\end{equation}

where the product refers to matrix multiplication. Swap operators for
qubits \(a\) and \(b\) can be applied to a density matrix
\(\rho\) in an \(N\) qubit system representing arbitrary mixed states to produce \(\rho_{swapped}\)
given as 
\begin{equation} \label{eq: swapped_state} \rho_{swapped} = S^N_{ab}\rho (S^N_{ab})^{\dagger}\end{equation} .

In the case of three qubits, then, we would need to ensure the two qubit subsystems can reach the maximally mixed state, so we would need to create four independently varying pure states for one permutation. We also want to add together different permutations of state with different qubits entangle which can also vary freely, meaning we would need three sets of four pure states in the end, or 12 two qubit states and 12 one qubit states.

So for three qubits the value of \(A\) in Equation \ref{eq: mixed} is 12. Each two qubit state would need variables for 4 magnitudes and three phases, and each one qubit state would need variables for two magnitudes and one phase. We would also need 12 parameters to determine the level of mixing. In total, parameterizing a three qubit state would require \(12 \times (4 + 3 + 12)\times (2 + 1) + 12 = 132\) floating point variables to be handled by Tensorflow.

For 4 qubits, all permutations of the states of the form
\(12|34\)
and
\(1|234\)
were considered; so each possible permutation of two pairs of entangled qubits or one group of three entangled qubits. Similarly for 5 qubits, all permutations of states of the form
\(1|2345\), \(12|345\) and \(12|34|5\)
were considered; that is, every arrangement 5 qubits comprised of arbitrary 1, 2 and 3 qubit states. Note that different ordering of qubits which may be
entangled together do not comprise additional permutations- so for
example we consider \(12|34\) to be identical to \(21|43\), but \(13|24\) is different.
In total there were 7 distinct 4 qubit permutations and 30 distinct 5
qubit permutations, all of which are listed in Appendix \ref{sec:perms} with the corresponding swap operator(s) \( S_{ab}\) required to produce them.

In total to parameterize four qubit states, we have to consider three permutations of states with two groups of two qubits, and we need four independent states for each permutation to achieve the maximally mixed state. This amount to 12 pure states for this partition. For states with one qubit separate from the other three, we have to consider four permutations, and we must guarantee the three qubit subsystems can be maximally mixed, which means we need eight states for each permutation to cover all eight basis states. This amounts to eight states for each of four permutations, or 32 states in total. So for four qubits, the value of \(A\) in equation \ref{eq: mixed} is 44. Furthermore, as two qubit states need four magnitudes and three phases each, the states defined by partitions into two qubit groups need \(2 \times 12 \times (4+3) = 168\) floating point parameters. Since the three qubit states need eight magnitudes and seven phases, and single qubit states need two magnitudes and one phase, then the partitions of three qubits separable from one would need \(32 \times ((8+7) + (2+1)) = 576\) parameters. With an additional 44 parameters needed to account for mixing, four qubit mixed k-separable states need \(168 + 576 + 44 = 788\) floating point parameters in total. Similar reasoning can be used to show that for five qubits, \(A = 220\) in equation \ref{eq: mixed} and in total 5500 floating point parameters must be managed by Tensorflow.

Comparing the number of parameters we use to the number necessary to represent arbitrary three, four, and five qubit states, we can show another measure of redundancy which indicates we should be able too reach arbitrary k-separable mixed states. Since every k-separable mixed state can be represented by a single density matrix, we must have at least as many parameters in our representation as would be required in general.

Density matrices are Hermitian, and have a trace of 1, so for an \(n \times n\) density matrix, \(\frac{(n)(n+1)}{2}\) real parameters are required (the lower triangle and the diagonal), and \(\frac{(n)(n-1)}{2}\) imaginary parameters are required (the lower triangle only). So a three qubit density matrix in general should require no more than 64 parameters, while we use 132. A four qubit matrix would require at least 256 parameters, while we use 788 in total, and a five qubit density matrix would need 1024, while we use 5500.

\section{\label{sec:perms} Permutations of 4 and 5 Qubit Pure K-Separable States}

\begin{table}[h]
\caption{4 Qubit Permutations of 1\textbar234}
\begin{ruledtabular}
\begin{tabular}{c c c c}
\textbf{Qubit Arrangement} & \textbf{Swap Operators}\\
\hline
\textbf{1\textbar234} & None \\
\textbf{2\textbar134} & \(S_{12}^4\) \\
\textbf{3\textbar124} & \(S_{13}^4\) \\
\textbf{4\textbar123} & \(S_{14}^4\) \\
\end{tabular}
\end{ruledtabular}
\end{table}

\begin{table}[h]
\caption{\label{tab:table3} 4 Qubit Permutations of 12\textbar34}
\begin{ruledtabular}
\begin{tabular}{c c c c}
\textbf{Qubit Arrangement} & \textbf{Swap Operators}\\
\hline
\textbf{12\textbar34} & None \\
\textbf{13\textbar24} & \(S_{23}^4\) \\
\textbf{14\textbar23} & \(S_{24}^4\) \\
\end{tabular}
\end{ruledtabular}
\end{table}

\begin{table}[h]
\caption{\label{tab:table4} 5 Qubit Permutations of 1\textbar2345}
\begin{ruledtabular}
\begin{tabular}{c c c c}
\textbf{Qubit Arrangement} & \textbf{Swap Operators}\\
\hline
\textbf{1\textbar2345} & None \\
\textbf{2\textbar1345} & \(S_{12}^5\) \\
\textbf{3\textbar1245} & \(S_{13}^5\) \\
\textbf{4\textbar1235} & \(S_{14}^5\) \\
\textbf{5\textbar1234} & \(S_{15}^5\) \\
\end{tabular}
\end{ruledtabular}
\end{table}

\begin{table}[h]
\caption{\label{tab:table5} 5 Qubit Permutations of 12\textbar345}
\begin{ruledtabular}
\begin{tabular}{c c c c}
\textbf{Qubit Arrangement} & \textbf{Swap Operators}\\
\hline
\textbf{12\textbar345} & None \\
\textbf{13\textbar245} & \(S_{23}^5\) \\
\textbf{14\textbar325} & \(S_{24}^5\) \\
\textbf{15\textbar324} & \(S_{25}^5\) \\
\textbf{23\textbar145} & \(S_{13}^5\) \\
\textbf{24\textbar135} & \(S_{14}^5\) \\
\textbf{25\textbar134} & \(S_{15}^5\) \\
\textbf{35\textbar124} & \(S_{23}^5S_{15}^5\) \\
\textbf{45\textbar123} & \(S_{24}^5S_{15}^5\) \\
\textbf{34\textbar125} & \(S_{23}^5S_{14}^5\) \\
\end{tabular}
\end{ruledtabular}
\end{table}

\begin{table}[h]
\caption{\label{tab:table6} 5 Qubit Permutations of 12\textbar34\textbar5}
\begin{ruledtabular}
\begin{tabular}{c c c c}
\textbf{Qubit Arrangement} & \textbf{Swap Operators}\\
\hline
\textbf{12\textbar34\textbar5} & None \\
\textbf{13\textbar24\textbar5} & \(S_{23}^5\) \\
\textbf{14\textbar32\textbar5} & \(S_{24}^5\) \\
\textbf{15\textbar34\textbar2} & \(S_{25}^5\) \\
\textbf{25\textbar34\textbar1} & \(S_{15}^5\) \\
\textbf{12\textbar45\textbar3} & \(S_{35}^5\) \\
\textbf{12\textbar35\textbar4} & \(S_{45}^5\) \\
\textbf{35\textbar24\textbar1} & \(S_{23}^5S_{15}^5\) \\
\textbf{45\textbar23\textbar1} & \(S_{24}^5S_{15}^5\) \\
\textbf{35\textbar14\textbar2} & \(S_{25}^5S_{31}^5\) \\
\textbf{45\textbar13\textbar2} & \(S_{25}^5S_{41}^5\) \\
\textbf{14\textbar25\textbar3} & \(S_{35}^5S_{24}^5\) \\
\textbf{24\textbar15\textbar3} & \(S_{35}^5S_{14}^5\) \\
\textbf{13\textbar25\textbar4} & \(S_{45}^5S_{23}^5\) \\
\textbf{23\textbar15\textbar4} & \(S_{45}^5S_{13}^5\) \\
\end{tabular}
\end{ruledtabular}
\end{table}

\section{\label{sec:coeff} 4 and 5 Qubit W State Witness Coefficients}

\begin{table}[h]
\caption{\label{tab: W witness, 46 terms} 46 Term 4 Qubit W State Witness Coefficients}
\begin{ruledtabular}
\begin{tabular}{c c c c}
\textbf{Feature} & \textbf{Coefficient Value}\\
\hline
\textbf{IIII} & 1.0 \\
\textbf{IIIZ} & 0.0419743592 \\
\textbf{IIZI} & 0.0407342861 \\
\textbf{IXXZ} & -0.0432833501 \\
\textbf{IXZX} & -0.0471624043 \\
\textbf{IYYZ} & -0.0536698235 \\
\textbf{IYZY} & -0.0530854546 \\
\textbf{IZII} & 0.040525885 \\
\textbf{IZXX} & -0.0437433997 \\
\textbf{IZYY} & -0.0543628183 \\
\textbf{IZZZ} & 0.0421243215 \\
\textbf{XIXZ} & -0.0478512441 \\
\textbf{XIZX} & -0.0464405696 \\
\textbf{XXIZ} & -0.0464631243 \\
\textbf{XXZI} & -0.0471208157 \\
\textbf{XXZZ} & -0.1161227503 \\
\textbf{XZIX} & -0.0469028852 \\
\textbf{XZXI} & -0.0494780751 \\
\textbf{XZXZ} & -0.1142446161 \\
\textbf{XZZX} & -0.1153172026 \\
\textbf{YIYZ} & -0.0548238079 \\
\textbf{YIZY} & -0.0548713095 \\
\textbf{YYIZ} & -0.0554185521 \\
\textbf{YYZI} & -0.0558558046 \\
\textbf{YYZZ} & -0.1145918067 \\
\textbf{YZIY} & -0.0546492017 \\
\textbf{YZYI} & -0.0563558375 \\
\textbf{YZYZ} & -0.1138392217 \\
\textbf{YZZY} & -0.1143895459 \\
\textbf{ZIII} & 0.0403184898 \\
\textbf{ZIXX} & -0.0441186353 \\
\textbf{ZIYY} & -0.0549539893 \\
\textbf{ZIZZ} & 0.0413865868 \\
\textbf{ZXIX} & -0.0475253985 \\
\textbf{ZXXI} & -0.0446825172 \\
\textbf{ZXXZ} & -0.1141413184 \\
\textbf{ZXZX} & -0.1155061694 \\
\textbf{ZYIY} & -0.0535187582 \\
\textbf{ZYYI} & -0.055262925 \\
\textbf{ZYYZ} & -0.1150405884 \\
\textbf{ZYZY} & -0.1152323382 \\
\textbf{ZZIZ} & 0.0435408697 \\
\textbf{ZZXX} & -0.1146970388 \\
\textbf{ZZYY} & -0.1147030555 \\
\textbf{ZZZI} & 0.0431077432 \\
\textbf{ZZZZ} & 0.1332015779 \\
\end{tabular}
\end{ruledtabular}
\end{table}

\begin{table}[h]
\caption{\label{tab: W witness, 38 terms} 38 Term 4 Qubit W State Witness Coefficients}
\begin{ruledtabular}
\begin{tabular}{c c c c}
\textbf{Feature} & \textbf{Coefficient Value}\\
\hline
\textbf{IIII} & 1.0 \\
\textbf{IIIZ} & 0.0333671311 \\
\textbf{IIZI} & 0.0779614398 \\
\textbf{IXZX} & -0.0508396477 \\
\textbf{IYYZ} & -0.0667737908 \\
\textbf{IYZY} & -0.0669929573 \\
\textbf{IZII} & 0.0792513558 \\
\textbf{IZXX} & -0.0579823307 \\
\textbf{IZYY} & -0.0721423199 \\
\textbf{XIXZ} & -0.0827026619 \\
\textbf{XIZX} & -0.0856182584 \\
\textbf{XXIZ} & -0.0827892765 \\
\textbf{XXZI} & -0.0429017171 \\
\textbf{XXZZ} & -0.1917474616 \\
\textbf{XZIX} & -0.0830169549 \\
\textbf{XZXI} & -0.0511875355 \\
\textbf{XZXZ} & -0.1843460324 \\
\textbf{XZZX} & -0.1959787667 \\
\textbf{YIYZ} & -0.0838920882 \\
\textbf{YIZY} & -0.0973450315 \\
\textbf{YYIZ} & -0.0825498141 \\
\textbf{YYZI} & -0.0645663378 \\
\textbf{YYZZ} & -0.1775653202 \\
\textbf{YZIY} & -0.1008676592 \\
\textbf{YZYI} & -0.0638608135 \\
\textbf{YZYZ} & -0.1779474813 \\
\textbf{YZZY} & -0.1708651555 \\
\textbf{ZIII} & 0.0391598295 \\
\textbf{ZIXX} & -0.0830765202 \\
\textbf{ZIYY} & -0.0828133173 \\
\textbf{ZXIX} & -0.0825274139 \\
\textbf{ZXZX} & -0.1854921375 \\
\textbf{ZYIY} & -0.0817359589 \\
\textbf{ZYYI} & -0.0675960989 \\
\textbf{ZYYZ} & -0.1657540624 \\
\textbf{ZYZY} & -0.1760615175 \\
\textbf{ZZXX} & -0.1807754956 \\
\textbf{ZZYY} & -0.1731762291 \\
\end{tabular}
\end{ruledtabular}
\end{table}

\begin{table}[h]
\caption{\label{tab: W witness 28 terms} 28 Term 4 Qubit W State Witness Coefficients}
\begin{ruledtabular}
\begin{tabular}{c c c c}
\textbf{Feature} & \textbf{Coefficient Value}\\
\hline
\textbf{IIII} & 1.0 \\
\textbf{IXZX} & -0.050627399 \\
\textbf{IZII} & 0.1428474669 \\
\textbf{IZXX} & -0.0729758589 \\
\textbf{IZYY} & -0.0725457014 \\
\textbf{XIXZ} & -0.1646973039 \\
\textbf{XIZX} & -0.1526620547 \\
\textbf{XXZZ} & -0.2500445496 \\
\textbf{XZIX} & -0.0576395294 \\
\textbf{XZXI} & -0.0712236378 \\
\textbf{XZXZ} & -0.2035209395 \\
\textbf{XZZX} & -0.2301877977 \\
\textbf{YIYZ} & -0.1446993769 \\
\textbf{YIZY} & -0.1446396575 \\
\textbf{YYIZ} & -0.1368158313 \\
\textbf{YYZI} & -0.133058841 \\
\textbf{YZIY} & -0.0855100987 \\
\textbf{YZYI} & -0.0753752536 \\
\textbf{YZYZ} & -0.2019849266 \\
\textbf{YZZY} & -0.197404676 \\
\textbf{ZIXX} & -0.1664206334 \\
\textbf{ZIYY} & -0.1437308329 \\
\textbf{ZXIX} & -0.0631352414 \\
\textbf{ZXZX} & -0.224047375 \\
\textbf{ZYYI} & -0.1420445032 \\
\textbf{ZYZY} & -0.2717156322 \\
\textbf{ZZXX} & -0.2258905471 \\
\textbf{ZZYY} & -0.2108470064 \\
\end{tabular}
\end{ruledtabular}
\end{table}

\begin{table}[h]
\caption{\label{tab: W witness, 20 terms} 20 Term 5 Qubit W State Witness Coefficients}
\begin{ruledtabular}
\begin{tabular}{c c c c}
\textbf{Feature} & \textbf{Coefficient Value}\\
\hline
\textbf{IIIII} & 1.0 \\
\textbf{IZZZZ} & 0.0950641886 \\
\textbf{XIXZZ} & -0.1032873369 \\
\textbf{XXZZZ} & -0.0860556505 \\
\textbf{YYZZZ} & -0.113579264 \\
\textbf{YZYZZ} & -0.113756095 \\
\textbf{YZZYZ} & -0.1142755273 \\
\textbf{YZZZY} & -0.1139525742 \\
\textbf{ZIZXX} & -0.1022777402 \\
\textbf{ZIZZZ} & 0.0943556723 \\
\textbf{ZYYZZ} & -0.1142736966 \\
\textbf{ZYZYZ} & -0.1128594578 \\
\textbf{ZYZZY} & -0.1129540724 \\
\textbf{ZZIZZ} & 0.0902691677 \\
\textbf{ZZXZX} & -0.1082390016 \\
\textbf{ZZYYZ} & -0.1131057401 \\
\textbf{ZZZIZ} & 0.0930643682 \\
\textbf{ZZZYY} & -0.113770213 \\
\textbf{ZZZZI} & 0.091169149 \\
\textbf{ZZZZZ} & 0.2121704484 \\
\end{tabular}
\end{ruledtabular}
\end{table}

\begin{longtable}[B]{c c}
\caption{180 Term 5 Qubit W State Witness Coefficients} \label{tab: W witness, 180 terms} \\
\toprule
\textbf{Feature} & \textbf{Coefficient Value}\\
\hline
\textbf{IIIII} & 1.0 \\
\textbf{IIIIZ} & -0.0004849752 \\
\textbf{IIIXX} & -0.008102558 \\
\textbf{IIIYY} & -0.0054531029 \\
\textbf{IIXIX} & -0.0046863507 \\
\textbf{IIXXI} & -0.0075433074 \\
\textbf{IIXXZ} & -0.0024097769 \\
\textbf{IIXZX} & -0.0041934936 \\
\textbf{IIYIY} & -0.0038011316 \\
\textbf{IIYYI} & -0.0044843634 \\
\textbf{IIYYZ} & -0.003163367 \\
\textbf{IIYZY} & -0.0047493098 \\
\textbf{IIZII} & -0.0023932361 \\
\textbf{IIZIZ} & -0.0008611195 \\
\textbf{IIZXX} & -0.0055868119 \\
\textbf{IIZYY} & -0.0043207945 \\
\textbf{IIZZI} & -0.0007211991 \\
\textbf{IXIIX} & -0.0025229831 \\
\textbf{IXIXI} & -0.0082578863 \\
\textbf{IXIXZ} & -0.0051665217 \\
\textbf{IXIZI} & 0.0004781462 \\
\textbf{IXIZX} & -0.0044166116 \\
\textbf{IXXII} & -0.012479162 \\
\textbf{IXXIZ} & -0.0014192292 \\
\textbf{IXXZI} & -0.011719729 \\
\textbf{IXXZZ} & -0.0158728599 \\
\textbf{IXZIX} & -0.002313323 \\
\textbf{IXZXI} & -0.0040635108 \\
\textbf{IXZXZ} & -0.0207355856 \\
\textbf{IXZZX} & -0.0160843189 \\
\textbf{IYIIY} & -0.0040437635 \\
\textbf{IYIYI} & -0.0048844941 \\
\textbf{IYIYZ} & -0.0060971465 \\
\textbf{IYIZY} & -0.0041654083 \\
\textbf{IYYII} & -0.0065254659 \\
\textbf{IYYIZ} & -0.0066039092 \\
\textbf{IYYZI} & -0.0046088682 \\
\textbf{IYYZZ} & -0.0151188845 \\
\textbf{IYZIY} & -0.0051449427 \\
\textbf{IYZYI} & -0.0040005256 \\
\textbf{IYZYZ} & -0.0187702691 \\
\textbf{IYZZY} & -0.0178758525 \\
\textbf{IZIIX} & -0.0004747625 \\
\textbf{IZIXX} & -0.0023253412 \\
\textbf{IZIYY} & -0.0051187037 \\
\textbf{IZXIX} & -0.0036619507 \\
\textbf{IZXXI} & -0.0018159616 \\
\textbf{IZXXZ} & -0.0020687413 \\
\textbf{IZXZX} & -0.0146048751 \\
\textbf{IZYIY} & -0.0039438247 \\
\textbf{IZYYI} & -0.0042955883 \\
\textbf{IZYYZ} & -0.0158398088 \\
\textbf{IZYZY} & -0.0162150404 \\
\textbf{IZZII} & -0.0011404493 \\
\textbf{IZZXX} & -0.0135089298 \\
\textbf{IZZYY} & -0.0164221719 \\
\textbf{IZZZI} & 0.0005482441 \\
\textbf{IZZZX} & 0.0004682327 \\
\textbf{IZZZZ} & 0.0834496976 \\
\textbf{XIIIX} & -0.0075410248 \\
\textbf{XIIXI} & -0.0109499371 \\
\textbf{XIIXZ} & -0.0062930585 \\
\textbf{XIIZX} & -0.0063538044 \\
\textbf{XIXII} & -0.0075533749 \\
\textbf{XIXIZ} & -0.0014604978 \\
\textbf{XIXZI} & -0.0058500566 \\
\textbf{XIXZZ} & -0.0198717398 \\
\textbf{XIZIX} & -0.0038767749 \\
\textbf{XIZXI} & -0.0029974829 \\
\textbf{XIZXZ} & -0.0054102408 \\
\textbf{XIZZX} & -0.0084536028 \\
\textbf{XXIII} & -0.0077317525 \\
\textbf{XXIIZ} & -0.000667373 \\
\textbf{XXIZI} & -0.0037931514 \\
\textbf{XXIZZ} & -0.0025216733 \\
\textbf{XXZII} & -0.0112564262 \\
\textbf{XXZIZ} & -0.0115440899 \\
\textbf{XXZZI} & -0.0292754406 \\
\textbf{XXZZZ} & -0.0146743152 \\
\textbf{XZIIX} & -0.0057336375 \\
\textbf{XZIXI} & -0.0090726932 \\
\textbf{XZIXZ} & -0.0188265673 \\
\textbf{XZIZX} & -0.0184168171 \\
\textbf{XZXII} & -0.0044191075 \\
\textbf{XZXIZ} & -0.0165701955 \\
\textbf{XZXZI} & -0.0137146579 \\
\textbf{XZXZZ} & -0.0041604784 \\
\textbf{XZZIX} & -0.0130517107 \\
\textbf{XZZXI} & -0.0154965877 \\
\textbf{XZZXZ} & -0.0053352605 \\
\textbf{XZZZX} & -0.0157124417 \\
\textbf{YIIIY} & -0.0045976941 \\
\textbf{YIIYI} & -0.003823808 \\
\textbf{YIIYZ} & -0.0054076913 \\
\textbf{YIIZY} & -0.0048135411 \\
\textbf{YIYII} & -0.0064142622 \\
\textbf{YIYIZ} & -0.0056990956 \\
\textbf{YIYZI} & -0.0049174563 \\
\textbf{YIYZZ} & -0.0141448146 \\
\textbf{YIZIY} & -0.0064018468 \\
\textbf{YIZYI} & -0.0047559117 \\
\textbf{YIZYZ} & -0.0175711856 \\
\textbf{YIZZY} & -0.018384868 \\
\textbf{YYIII} & -0.0059897102 \\
\textbf{YYIIZ} & -0.0053822238 \\
\textbf{YYIZI} & -0.0061193625 \\
\textbf{YYIZZ} & -0.0169064427 \\
\textbf{YYZII} & -0.0049580991 \\
\textbf{YYZIZ} & -0.0157065482 \\
\textbf{YYZZI} & -0.0158290078 \\
\textbf{YYZZZ} & -0.0376947502 \\
\textbf{YZIIY} & -0.0043922613 \\
\textbf{YZIYI} & -0.0036898327 \\
\textbf{YZIYZ} & -0.01456578 \\
\textbf{YZIZY} & -0.0131398391 \\
\textbf{YZYII} & -0.0051867058 \\
\textbf{YZYIZ} & -0.0149939536 \\
\textbf{YZYZI} & -0.0148226813 \\
\textbf{YZYZZ} & -0.038268032 \\
\textbf{YZZIY} & -0.0162159378 \\
\textbf{YZZYI} & -0.0144634604 \\
\textbf{YZZYZ} & -0.0417575946 \\
\textbf{YZZZY} & -0.0405910077 \\
\textbf{ZIIXX} & -0.0075723394 \\
\textbf{ZIIYY} & -0.0043800479 \\
\textbf{ZIXIX} & -0.001250673 \\
\textbf{ZIXXI} & -0.0095401609 \\
\textbf{ZIXXZ} & -0.0238106857 \\
\textbf{ZIXZX} & -0.0180715461 \\
\textbf{ZIYIY} & -0.0042049701 \\
\textbf{ZIYYI} & -0.0056919552 \\
\textbf{ZIYYZ} & -0.0140640307 \\
\textbf{ZIYZY} & -0.0152214709 \\
\textbf{ZIZII} & -0.0010793222 \\
\textbf{ZIZXX} & -0.0226945714 \\
\textbf{ZIZYY} & -0.0145368859 \\
\textbf{ZIZZZ} & 0.0826497 \\
\textbf{ZXIIX} & -0.0013379113 \\
\textbf{ZXIXI} & -0.0022694798 \\
\textbf{ZXIXZ} & -0.0125266469 \\
\textbf{ZXIZX} & -0.0115735332 \\
\textbf{ZXXII} & -0.0077368178 \\
\textbf{ZXXIZ} & -0.0049935527 \\
\textbf{ZXXZI} & -0.0179745189 \\
\textbf{ZXXZZ} & -0.0020596257 \\
\textbf{ZXZIX} & -0.0168591727 \\
\textbf{ZXZXI} & -0.0121962055 \\
\textbf{ZXZXZ} & -0.006862098 \\
\textbf{ZXZZX} & -0.0213124501 \\
\textbf{ZYIIY} & -0.0040482449 \\
\textbf{ZYIYI} & -0.0031516294 \\
\textbf{ZYIYZ} & -0.0148053857 \\
\textbf{ZYIZY} & -0.0126354755 \\
\textbf{ZYYII} & -0.0062106044 \\
\textbf{ZYYIZ} & -0.0164441946 \\
\textbf{ZYYZI} & -0.0146876056 \\
\textbf{ZYYZZ} & -0.0373226252 \\
\textbf{ZYZIY} & -0.0147140033 \\
\textbf{ZYZYI} & -0.0109277099 \\
\textbf{ZYZYZ} & -0.0419482307 \\
\textbf{ZYZZY} & -0.0404860063 \\
\textbf{ZZIII} & 0.0007225496 \\
\textbf{ZZIXX} & -0.0042798907 \\
\textbf{ZZIYY} & -0.0150632424 \\
\textbf{ZZIZI} & 0.0010620551 \\
\textbf{ZZIZZ} & 0.0839926033 \\
\textbf{ZZXIX} & -0.0157195566 \\
\textbf{ZZXXI} & -0.0258583231 \\
\textbf{ZZXXZ} & -0.002876667 \\
\textbf{ZZXZX} & -0.0133240713 \\
\textbf{ZZYIY} & -0.0131715733 \\
\textbf{ZZYYI} & -0.0160673905 \\
\textbf{ZZYYZ} & -0.0424820497 \\
\textbf{ZZYZY} & -0.0418729995 \\
\textbf{ZZZIZ} & 0.0814709818 \\
\textbf{ZZZXX} & -0.0154490959 \\
\textbf{ZZZYY} & -0.0402438112 \\
\textbf{ZZZZI} & 0.082902013 \\
\textbf{ZZZZX} & 0.0004571153 \\
\textbf{ZZZZZ} & 0.1925381066 \\
\toprule
\end{longtable}

\end{document}